%% file: main.tex
\documentclass[sigconf, nonacm, pdfa]{acmart}
\usepackage[a-2b]{pdfx}
\usepackage{amsmath}
\usepackage{bm}
\usepackage{graphicx}
\usepackage{etoolbox}
\usepackage{enumitem}
\usepackage[noend]{algorithmic}
\usepackage{multirow}
\usepackage{amsthm}
\usepackage{xspace}
\usepackage{amsmath,stmaryrd,pict2e,picture}
\usepackage{amsmath}
\usepackage{bbold}
\usepackage{multirow}
\usepackage{diagbox}
\usepackage{balance}
\usepackage{float}
\usepackage[noend,lined,boxed,vlined,ruled,linesnumbered]
{algorithm2e}

\usepackage{caption, subcaption}

\usepackage[labelformat=simple]{subcaption}
\SetKwInOut{Input}{input}
\SetKwInOut{Output}{output}

\DeclareMathOperator*{\argmax}{arg\,max}

\newcommand{\code}[1]{\texttt{#1}}
\newcommand{\at}{\textsc{Auto-Tables}\xspace}
\newcommand{\atbench}{\textsc{ATBench}\xspace}

\newcounter{definition}
\newenvironment{definition}[1][]{\refstepcounter{definition}\par\smallskip\textsc{Definition~\thedefinition.\ #1}}{\smallskip}

\newcounter{example}
\newenvironment{example}[1][]{\refstepcounter{example}\par\smallskip\textsc{Example~\theexample.\ #1}}{\smallskip}

\newcounter{theorem}

\newcounter{proposition}

\newtoggle{fullversion}
\toggletrue{fullversion}
\newcommand\vldbdoi{10.14778/3611479.3611534}
\newcommand\vldbpages{3391 - 3403}
\newcommand\vldbvolume{16}
\newcommand\vldbissue{11}
\newcommand\vldbyear{2023}

\newcommand\vldbauthors{\authors}
\newcommand\vldbtitle{\shorttitle} 
\newcommand\vldbavailabilityurl{https://github.com/LiPengCS/Auto-Tables-Benchmark}
\newcommand\vldbpagestyle{empty} 
\newcommand{\stitle}[1]{\vspace{1ex}\noindent{\bf #1}}

\begin{document}
\title{Auto-Tables: Synthesizing Multi-Step Transformations to Relationalize Tables without Using Examples}

%
% The "author" command and its associated commands are used to define the authors and their affiliations.
\author{Peng Li}
\authornote{Part of work done while at Microsoft.}
\affiliation{%
  \institution{Georgia Tech}
}
\email{pengli@gatech.edu}

% \author{Yeye He, Cong Yan, Yue Wang, Surajit Chaudhuri}
% \affiliation{%
%   \institution{Microsoft Research}
%   %\streetaddress{1 Th{\o}rv{\"a}ld Circle}
%   %\city{Hekla}
%   %\country{Iceland}
% }
% \email{{yeyehe, coyan, wanyue, surajitc}@microsoft.com}

\author{Yeye He}
\affiliation{%
  \institution{Microsoft Research}
  %\streetaddress{1 Th{\o}rv{\"a}ld Circle}
  %\city{Hekla}
  %\country{Iceland}
}
\email{yeyehe@microsoft.com}

\author{Cong Yan}
\affiliation{%
  \institution{Microsoft Research}
  %\streetaddress{1 Th{\o}rv{\"a}ld Circle}
  %\city{Hekla}
  %\country{Iceland}
}
\email{coyan@microsoft.com}

\author{Yue Wang}
\affiliation{%
  \institution{Microsoft Research}
  %\streetaddress{1 Th{\o}rv{\"a}ld Circle}
  %\city{Hekla}
  %\country{Iceland}
}
\email{wanyue@microsoft.com}

\author{Surajit Chaudhuri}
\affiliation{%
  \institution{Microsoft Research}
  %\streetaddress{1 Th{\o}rv{\"a}ld Circle}
  %\city{Hekla}
  %\country{Iceland}
}
\email{surajitc@microsoft.com}

\begin{abstract}
Relational tables, where each row corresponds to an entity and each column corresponds to an attribute, have been the standard for tables in relational databases. However, such a standard cannot be taken for granted when dealing with tables ``in the wild''. Our survey of real spreadsheet-tables and web-tables shows that over 30\% of such tables do not conform to the relational standard, for which complex table-restructuring transformations are needed before these tables can be queried easily using SQL-based tools. %(e.g., queried using SQL or spreadsheet interfaces). 
Unfortunately, the required transformations are non-trivial to program, which has become a substantial pain point for technical and non-technical users alike, as evidenced by large numbers of forum questions in places like StackOverflow and Excel/Tableau forums.

We develop an \at system that can automatically synthesize pipelines with multi-step transformations (in Python or other languages), to transform non-relational tables into standard relational forms for downstream analytics, obviating the need for users to manually program transformations. We compile an extensive benchmark for this new task, by collecting 244 real test cases from user spreadsheets and online forums.  Our evaluation suggests that \at can successfully synthesize transformations for over 70\% of test cases at interactive speeds, without requiring any input from users, making this an effective tool for both technical and non-technical users to prepare data for analytics.
%for a large fraction of real tables without requiring any input from users.
\end{abstract}

\maketitle

\pagestyle{\vldbpagestyle}
\begingroup\small\noindent\raggedright\textbf{PVLDB Reference Format:}\\
\vldbauthors. \vldbtitle. PVLDB, \vldbvolume(\vldbissue): \vldbpages, \vldbyear.\\
\href{https://doi.org/\vldbdoi}{doi:\vldbdoi}
\endgroup
\begingroup
\renewcommand\thefootnote{}\footnote{\noindent
This work is licensed under the Creative Commons BY-NC-ND 4.0 International License. Visit \url{https://creativecommons.org/licenses/by-nc-nd/4.0/} to view a copy of this license. For any use beyond those covered by this license, obtain permission by emailing \href{mailto:info@vldb.org}{info@vldb.org}. Copyright is held by the owner/author(s). Publication rights licensed to the VLDB Endowment. \\
\raggedright Proceedings of the VLDB Endowment, Vol. \vldbvolume, No. \vldbissue\ %
ISSN 2150-8097. \\
\href{https://doi.org/\vldbdoi}{doi:\vldbdoi} \\
}\addtocounter{footnote}{-1}\endgroup

% % %%% do not modify the following VLDB block %%
% %%% VLDB block start %%%
% \ifdefempty{\vldbavailabilityurl}{}{
% \vspace{.3cm}
% \begingroup\small\noindent\raggedright\textbf{Artifact Availability:}\\
% The data and/or other artifacts have been made available at \url{\vldbavailabilityurl}.
% \endgroup
% }
% % %%% VLDB block end %%%

\input{Introduction}
\input{Related}
\input{Preliminary}
\input{Method}

\input{experiment}
\input{Conclusions}

\vspace{-1mm}
\begin{acks}
\vspace{-1mm}
We thank Dr. Kexin Rong and Dr. Xu Chu for their generous support and valuable feedback, as well as three anonymous VLDB reviewers for their helpful comments on our manuscript.
\end{acks}

\nocite{gao2018navigating}

%\clearpage

\clearpage

%\balance

\bibliographystyle{ACM-Reference-Format}
\bibliography{AutoTables}

% \iftoggle{fullversion}
% {
%     % removed for revision
%     \revised{}
%     \clearpage
%     \appendix
%     %\input{apx-additional-operators.tex}
% }
% {
% }

\end{document}

%% file: Introduction.tex
\section{Introduction}
\label{sec:intro}
Modern data analytics like SQL and BI are predicated on a standard format of relational tables, where each row corresponds to a distinct ``entity'', and each column corresponds to an ``attribute'' for the entities that contains homogeneous data-values. While such tables are de-facto standards in relational databases, such that as database people we may take this for granted, a significant fraction of tables ``in the wild'' actually fail to conform to such standards, making them considerably more difficult to query using SQL-based tools.

\begin{figure*}[t]
\vspace{-15mm}
    \centering    
    \includegraphics[width=\textwidth]{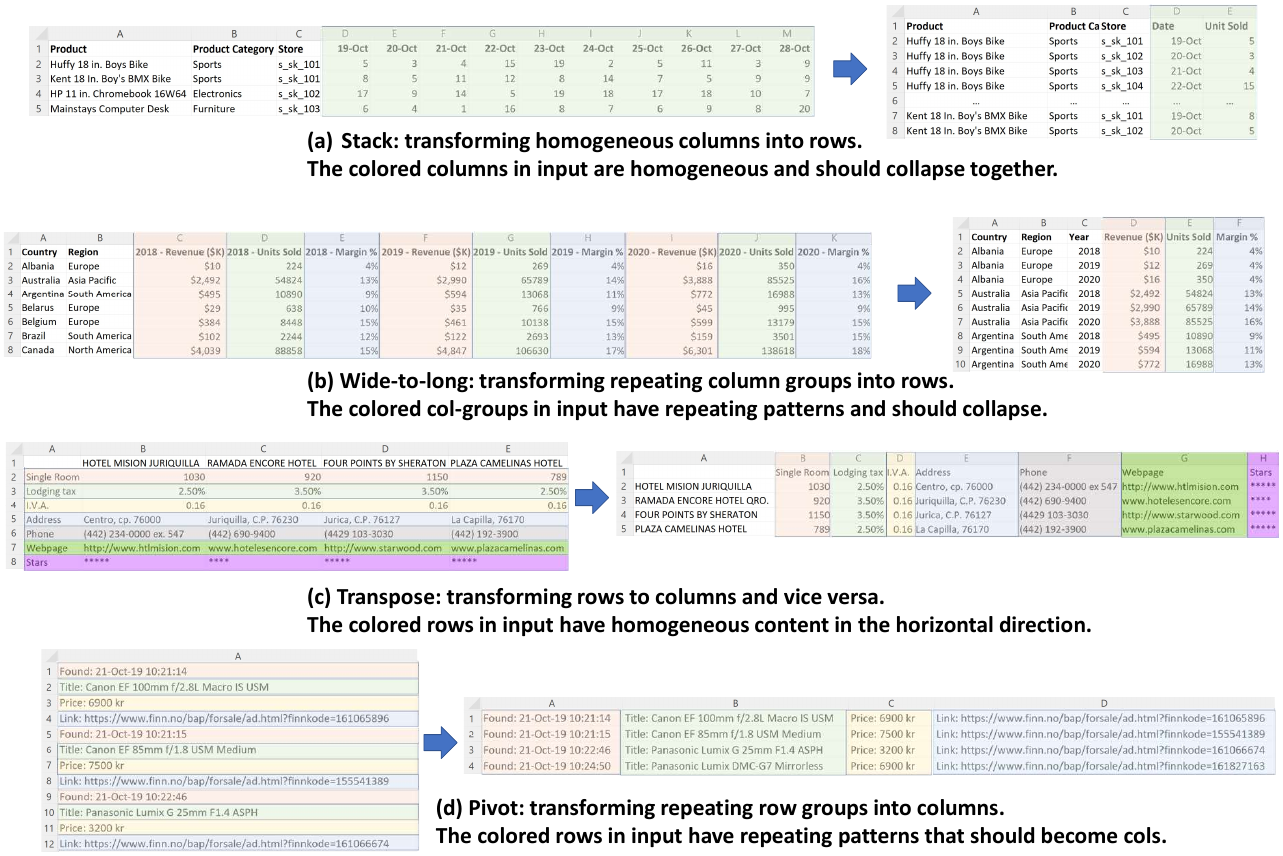}
    \vspace{-4mm}
    \caption{Example input/output tables for 4 operators in \at: (a) Stack, (b) Wide-to-long, (c) Transpose, (d) Pivot. The input-tables (on the left) are not relational and hard to query, which need to be transformed to produce corresponding output-tables (on the right) that are relational and easy to query. Observe that the color-coded, repeating row/column-groups are ``visual'' in nature, motivating a CNN-like architecture like used in computer vision for object-detection.}
    \label{fig:combined-ex}
\end{figure*}

\begin{figure*}[t]
    \centering    \includegraphics[width=2.0\columnwidth]{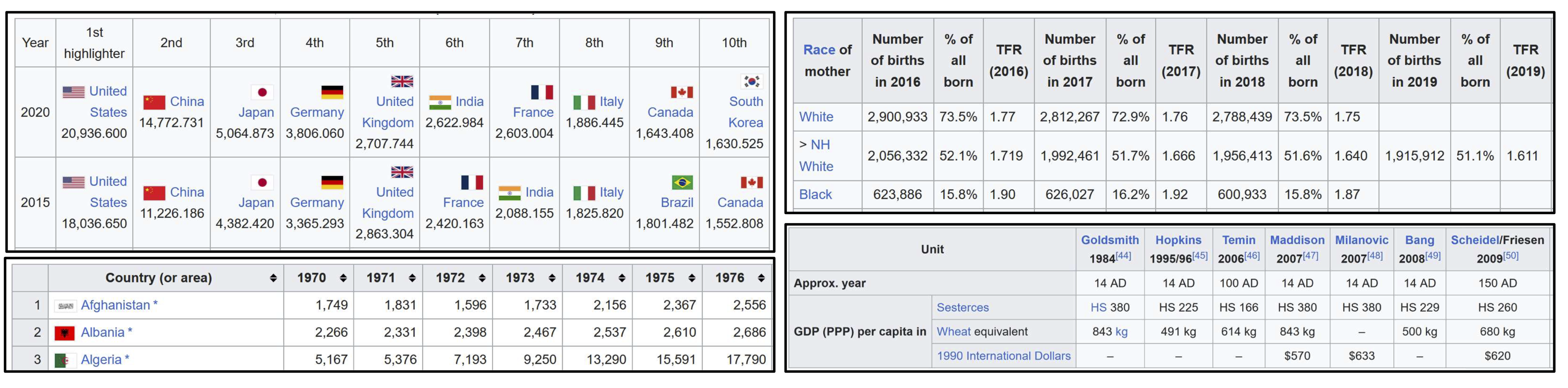}
    \vspace{-5mm}
    \caption{Real Web tables from Wikipedia that are also non-relational, similar to the spreadsheet tables shown in Figure~\ref{fig:combined-ex}.}
    \label{fig:combined-web-ex}
\end{figure*}

%web table examples:
% https://en.wikipedia.org/wiki/Demographics_of_the_United_States
% https://en.wikipedia.org/wiki/List_of_U.S._states_and_territories_by_historical_population
% https://en.wikipedia.org/wiki/List_of_countries_by_past_and_projected_GDP_(nominal)
% https://en.wikipedia.org/wiki/List_of_regions_by_past_GDP_(PPP)_per_capita
% https://en.wikipedia.org/wiki/List_of_countries_by_largest_historical_GDP

\textbf{Non-relational tables are common, but hard to query.} 
Real tables in the wild, such as spreadsheet-tables or web-tables, can often be ``non-relational'' and hard to query, unlike tables that we expect to find in relational databases. We randomly sampled hundreds of user spreadsheets (in Excel), and web tables (from Wikipedia), and found around 30-50\% tables to have such issues.
Figure~\ref{fig:combined-ex} and Figure~\ref{fig:combined-web-ex} show real samples taken
from spreadsheets and the web, respectively, to demonstrate these common issues. (We emphasize that the problem is prevalent at a very large scale, since there are millions of tables like these in spreadsheets and on the web.)

%and found these issues are widespread in real tables in the wild --  around 30-50\% tables we sampled have similar structural issues that make them hard to query. Considering that there are hundreds of millions of tables out there on the web and in spreadsheets, it is evident that the problem is prevalent on a large scale, if we want to help users unlock analytical value from these tables. %in analytical use cases.

%\yeye{add link to excel forum questions, to show non-relational tables cannot be queried easily}

%Figure~\ref{fig:combined-ex} and Figure~\ref{fig:combined-web-ex} show real examples taken from spreadsheets and the web, respectively.

%To illustrate common structural issues for these non-relational tables, we present four real examples taken from spreadsheets in the wild in Figure~\ref{fig:combined-ex}. 

Take Figure~\ref{fig:combined-ex}(a) for example. The table on the left is not a standard relational table, because each column marked in green contains sales numbers for only a single day (``\code{19-Oct}'', ``\code{20-Oct}'', etc.), making these column values highly homogeneous in the horizontal direction (while in typical relational tables, we expect values in columns to be homogeneous in the vertical direction). Although this specific table format makes it easy for humans to eyeball changes day-over-day by reading horizontally, it is unfortunately hard to analyze using SQL. Imagine that one needs to compute the 14-day average of sales, starting from  \code{``20-Oct''} -- for this table, one has to write: \code{SELECT SUM(``20-Oct'', ``21-Oct'', ``22-Oct'', ...) FROM T}, across 14 different columns, which is long and unwieldy to write. Now imagine we need 14-day moving averages with every day in October as the starting date -- the resulting SQL is highly repetitive and  hard to manage. %because these dates that are ordinarily data-values are now encoded as column-headers that  are hard to manipulate.

In contrast, consider a transformed version of this table, shown on the right of Figure~\ref{fig:combined-ex}(a). Here the homogeneous columns in the original table (marked in green) are transformed into only two new columns: ``\code{Date}'' and ``\code{Units Sold}'', using  a transformation operator called ``\code{stack}'' (listed in the first row of Table~\ref{tab:dsl}). This transformed table contains the same information as the original table, but is much easier to query -- e.g., the same 14-day moving average can be computed using a succinct range-predicate on the `\code{Date}'' column, where the starting date ``\code{20-Oct}'' is a literal parameter that can be easily changed into other values.

There are many such spreadsheet tables that require different kinds of transformations before they are ready for SQL-based analysis. Figure~\ref{fig:combined-ex}(b) shows another example where every 3 columns form a group, representing ``\code{Revenue}/\code{Units Sold}/\code{Margin}'' for a different year, repeating for many times (marked in red/green/blue in the figure). Tables with these repeating column-groups are also hard to query just like Figure~\ref{fig:combined-ex}(a), but in this case the required transformation operator is different and called ``\code{wide-to-long}'' (listed in the second row of Table~\ref{tab:dsl}).

Figure~\ref{fig:combined-ex}(c) shows yet another example, where each hotel corresponds to a column (whose names are in row-1), and each ``attribute'' of these hotels corresponds to a row. Note that in this case values in the same rows are homogeneous  (marked in different colors), unlike relational tables where values in the same columns are homogeneous. A transformation called ``\code{transpose}'' is required in this case (listed in the third row of Table~\ref{tab:dsl}), to make the resulting table, shown on the right of the figure, easy to query -- for instance, a query to sum up the total number of hotel rooms is hard to write on the original table, but can be easily achieved using a simple SUM query on the  ``\code{Single Room}'' column in the transformed table.

Figure~\ref{fig:combined-ex}(d) shows another example where columns are represented as rows in the table on the left. This is similar to Figure~\ref{fig:combined-ex}(c), except that the rows in this case are ``repeating'' in groups, thus requiring a different transformation operator called ``\code{pivot}'' (listed in the fourth row of Table~\ref{tab:dsl}) as opposed to ``\code{transpose}''.  The resulting table is shown on the right, which becomes easy to query.

While the examples so far are all taken from spreadsheets, we note that similar structural issues are also widespread in Web tables. Figure~\ref{fig:combined-web-ex} shows real examples from Wikipedia, which share similar characteristics as the spreadsheet tables in Figure~\ref{fig:combined-ex}, which all require transformations before these tables can be queried effectively. 

\begin{table*}[t]
\vspace{-15mm}
\caption{\at DSL: table-restructuring operators and their parameters to ``relationalize'' tables. These operators are common and exist in many different languages, like Python Pandas and R, sometimes under different names.}
\label{tab:dsl}
\vspace{-4mm}
\scalebox{0.8}{
\vspace{-16mm}
\hspace{-6mm}
\begin{tabular}{l|l|l|l}
\toprule
DSL operator & Python Pandas equivalent & Operator parameters & Description (example in parenthesis) \\
\midrule
stack & melt~\cite{op-melt} &  start\_idx, end\_idx & collapse homogeneous cols into rows (Fig.~\ref{fig:combined-ex}a) \\
wide-to-long & wide\_to\_long~\cite{op-wide-to-long} &  start\_idx, end\_idx, delim &  collapse repeating col-groups into rows (Fig.~\ref{fig:combined-ex}b) \\
transpose & transpose~\cite{op-transpose} & - & convert rows to columns and vice versa (Fig.~\ref{fig:combined-ex}c) \\
pivot & pivot~\cite{op-pivot} & repeat\_frequency & pivot repeating row-groups into cols (Fig.~\ref{fig:combined-ex}d) \\
explode & explode~\cite{op-explode} &  column\_idx, delim & convert composite cells into atomic values \\
ffill & ffill~\cite{op-ffill} & start\_idx, end\_idx & fill structurally empty cells in tables \\
subtitles & copy, ffill, del & column\_idx, row\_filter & convert table subtitles into a column \\
none &  - & -  & no-op, the input table is already relational \\
\bottomrule
\end{tabular}
}
\end{table*}

\textbf{Non-relational tables are hard to ``relationalize''.}
We mentioned that the example tables in Figure~\ref{fig:combined-ex} and Figure~\ref{fig:combined-web-ex} require different transformation operators. Table~\ref{tab:dsl} shows 8 such transformation operators commonly needed to relationalize tables (where the first 4 operators correspond to  the examples we see in Figure~\ref{fig:combined-ex}). 

The first column of Table~\ref{tab:dsl} shows the name of the ``logical operator'', which may be instantiated differently in different languages (e.g., in Python or R), with different names and syntax. The second column of the table shows the equivalent Pandas operator in Python~\cite{pandas}, which is a popular API for manipulating tables among developers and data scientists, that readers may be familiar with.
%. (Similar equivalent operators can be found in other languages such as R, with a different syntax.)

While the functionalities listed in Table~\ref{tab:dsl} already exist in languages such as R and Python, they are not easy for users to invoke correctly, because users need to:
\begin{enumerate}[leftmargin=*,noitemsep]
\item Visually identify different structural issues in an input table that make it hard to query (e.g., repeating row-/column-groups shown in Fig.~\ref{fig:combined-ex}(a-d)), which is not obvious to non-expert users; 
\item Map the visual pattern identified from the input table, to a corresponding operator in Table~\ref{tab:dsl} that can handle such issues. This is hard as users are often unfamiliar with the exact terminologies to describe these transformation operators (e.g., \code{pivot} vs. \code{stack}), often needing to search online for help;
\item Parameterize the chosen operator appropriately, using parameters tailored to the input table (e.g., which columns need to collapse into rows, what is the repeating frequency of column groups, etc.). This is again hard, as even developers need to consult the API documentation, which is often long and complex.
\item Certain input tables require more than one transformation step, for which users need to repeat steps (1)-(3) multiple times.
\end{enumerate}

\begin{figure}[t]
    \centering
    \includegraphics[width=0.7\columnwidth]{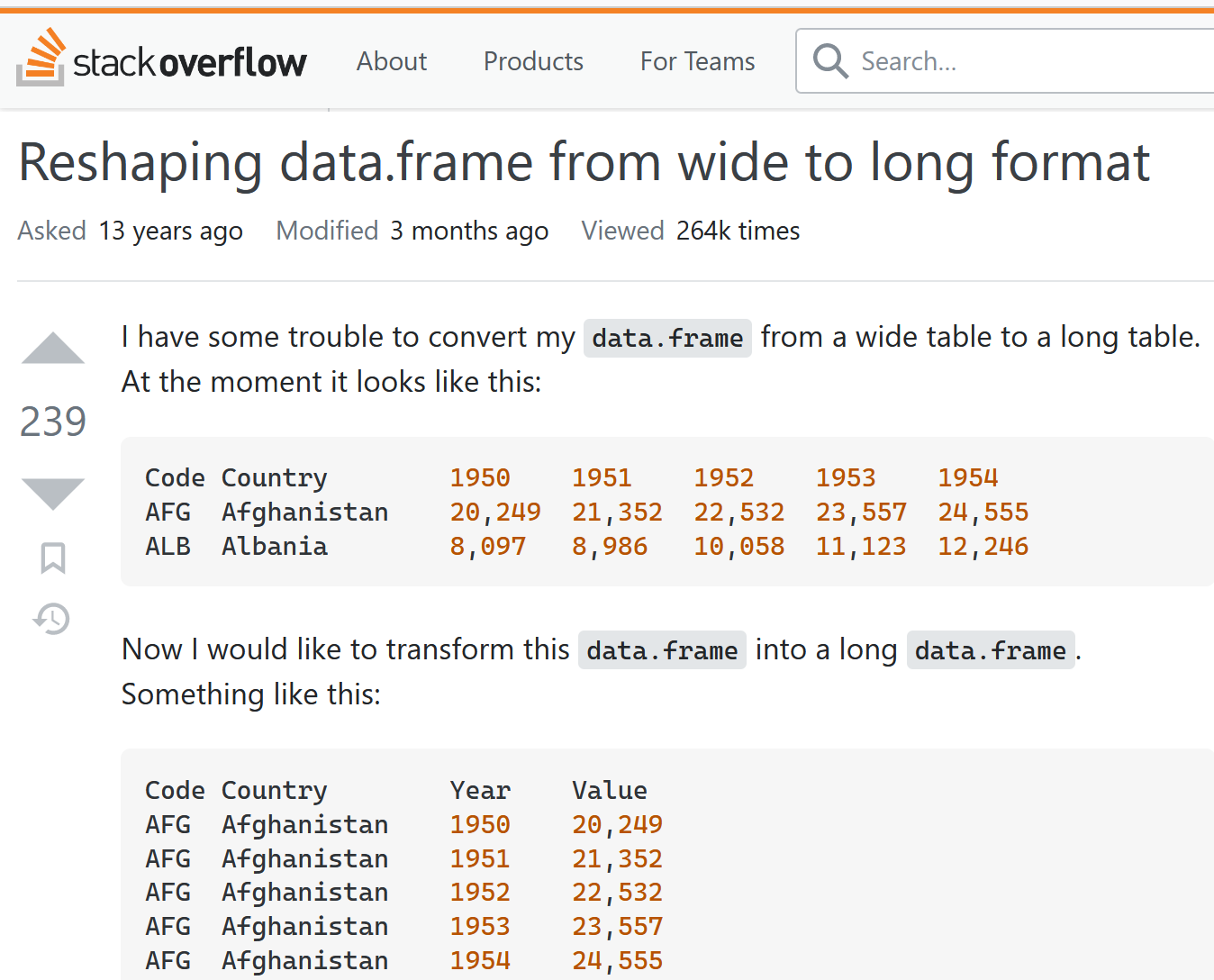}
    \vspace{-4mm}
    \caption{Example user question from StackOverflow, on how to restructure tables. Questions like this are common not only among technical users, but also non-technical users, as similar questions are commonly found on forums for Excel, Power-BI, and Tableau users too~\cite{excel-forum-1, excel-forum-2, excel-forum-3, excel-forum-4}.}
\vspace{-4mm}
\label{fig:stackoverflow-ex}
\end{figure}

Completing these steps is a tall order even for technical users, as evidenced by a large number of related questions on forums like StackOverflow (e.g.,~\cite{stackoverflow-forum-1, stackoverflow-forum-2, stackoverflow-forum-3, stackoverflow-forum-4}).  Figure~\ref{fig:stackoverflow-ex} shows such an example question (popular with many up-votes), where the developer provides example input/output tables to demonstrate the desired transformation, and seek help on what Pandas operators to invoke. %Note that questions like this are highly up-voted, showing that such questions are common pain points for developers. % (e.g., which Pandas API to invoke, and what parameters to use, etc.). Completing these steps is all the more challenging, for an average non-technical user who often need to manipulate tables in spreadsheets.
%(the authors to this date still need to consult the Pandas API reference pages of these operators to program them appropriately). 
%where developers ask for suggestions on how to restructure their tables. 

If technical users like developers find it hard to restructure their tables, as these StackOverflow questions would show, it comes as no surprise that non-technical enterprise users, who often deal with tables in spreadsheets, would find the task even more challenging. We find a large number of similar questions on Excel and Tableau forums (e.g.,~\cite{excel-forum-1, excel-forum-2, excel-forum-3, excel-forum-4}), where users complain that without the required transformations it is hard to analyze data using SQL-based or Excel-based tools (e.g.,~\cite{forum-hard-to-query-1, forum-hard-to-query-2, forum-hard-to-query-3, forum-hard-to-query-4}).

The prevalence of these  questions confirms table-restructuring as a common pain point for both technical and non-technical users.

\textbf{\at: synthesize transformations without examples.}
In this work, we propose a new paradigm to automatically synthesize table-restructuring steps to relationalize tables, using the Domain Specific Language (DSL) of operators in Table~\ref{tab:dsl}, \textit{without requiring users to provide examples}. Our key intuition of why we can do away with examples in our task, lies in the observation that given an input table, the logical steps required to relationalize it are \textit{almost always unique} and with little ambiguity, as the examples in Figure~\ref{fig:combined-ex} would all show. This is because the transformations required in our task only ``restructure'' tables, that do not actually ``alter'' the table content, which is unlike prior work that focuses on \textit{row-to-row transformations} (e.g., TDE~\cite{tde} and FlashFill~\cite{flashfill}), or SQL-by-example (e.g.~\cite{SQLSynthesizer, Scythe-code}), where the output is ``altered'' that can produce many possible outcomes, which would require users to provide input/output examples to demonstrate the desired outcome. %Asking for even a small number of output samples in our table-to-table transformation task means that users need to specify \textit{an entire output table}, which is a significant amount of effort, especially for large and complex real tables. 

%which typically require only 1-3 output examples, applying the same user-paradigm to our task of table-restructuring would mean that users have to specify \textit{an output table}, which is usually too much effort, especially for large and complex real tables.

%Our unique insight is that for our task, it may be possible to synthesize transformations without asking users to provide explicit examples. This is because to ``relationalize'' a given input table, the sequence of steps is almost always uniquely determined, given the unique characteristics of an input table as shown in Figure~\ref{fig:combined-ex}. 
For our task, we believe it is actually important \textit{not} to ask users to provide examples, because in the context of table-to-table transformations like in our case, asking users to provide examples would mean users have to specify \textit{an output table}, which is a substantial amount of typing effort, making it cumbersome to use.

%The reason we believe we can pull this off using algorithms and without examples, is because 
As humans, we can ``visually'' recognize  rows/columns patterns (e.g., homogeneous value groups, as color-coded vertically and horizontally in Figure~\ref{fig:combined-ex}), to correctly predict which operator to use. The question we ask in this paper, is whether an algorithm can ``learn'' to recognize such patterns by scanning the input tables alone, to predict suitable transformations, in a manner that is analogous to how computer-vision algorithms would scan a picture to identify common but more complex objects like dogs and cats.  

We should note that like computer vision problems such as object detection, where hand-crafted heuristics are hard to write, the row/column-level patterns existing in our target tables are also data-dependent and subtle, which are hard to write as heuristic rules. Consider for example the table in Figure~\ref{fig:combined-ex}(b) -- for ease of illustration we pick a case with three distinct groups of columns (currency, integers, and percentage-numbers, marked in different colors). One may hand-craft a heuristic ``similarity function'' between columns that may work for this simple example,  but imagine the common scenario where all these  columns have similar-looking integer numbers (e.g., with no dollar signs and percentage signs), which is much more challenging to predict using heuristics, as fine-grained differentiation is required to tell the subtle differences between columns (e.g., difference in column header semantics or column value ranges), which is best learned from the data. In fact, we tested a baseline using heuristic rules to predict only the simple ``stack'', which has a low 0.38 accuracy, because of the subtle differences in data that are not captured by heuristics. We also tested an LLM-based approach using GPT-3.5, also without success (with more details in our experiments), further underlining the challenging nature of our task. These motivate us to develop a learning-based method specifically tailored to our table transformation task.

In computer vision, in order to pick up subtle clues from pictures, object detection algorithms are typically trained using large amounts of labeled data~\cite{imagenet} (e.g., pictures of dogs that are manually labeled as such). In our task, we do not have such labeled datasets. Therefore, we devise a novel \textit{self-training framework} that exploits the \textit{inverse functional relationships} between operators (e.g., the inverse of ``\code{stack}'' is known as ``\code{unstack}''),  to automatically build large amounts of training data without requiring humans to label, as we will explain in Figure~\ref{fig:inverse-function}. Briefly, in order to build a training example for operator $O$ (e.g., ``\code{stack}''), we start from a relational table $R$ and apply the inverse of $O$, denoted by $O^{-1}$ (e.g., ``\code{unstack}''), to generate a table $T = O^{-1}(R)$, which we know is non-relational. For our task, given $T$ as input, we know $O$ must be its ground-truth transformation, because by definition $O(T) = O(O^{-1}(R))=R$, which turns $T$ back to its relational form $R$. This makes $(T, O)$ an (example, label) pair that we can automatically generate at scale, and use as our training data. 

%This is possible because starting from a relational table $R$, and in order to build a training example for operator $O$ (e.g., ``\code{stack}''), we can first apply the inverse of $O$, denoted by $O^{-1}$ (e.g., ``\code{unstack}''), to generate a table $T = O^{-1}(R)$. 

Leveraging training data so generated, we  develop an \at system that can ``learn-to-synthesize'' table-restructuring transformations, using a deep tabular model we develop inspired by CNN-like architectures popular in the computer vision literature. We show our approach is effective on real-world tasks, which can solve over 70\% of test cases collected from user forums and spreadsheets, while being interactive with sub-second latency.

%- where to get training? leverage inverse relationships of operators to learn from self-supervision.

%- show an example output program in python

%- why existing work is insufficient (by-example, too cumbersome)

\textbf{Contributions.} We make these contributions in \at:

\begin{itemize}[noitemsep,topsep=0pt,leftmargin=*]
\item We propose a novel problem to automatically relationalize tables without examples, which addresses a common pain point for both technical and non-technical users, when they deal with tables in the wild outside of database settings.
\item We develop \at that learns-to-synthesize transformations, using a computer-vision inspired model architecture that exploits the common ``visual'' patterns in tables.
\item We propose a self-supervision framework unique in our setting to overcome the lack of training data, by exploiting the inverse functional relationships between operators to auto-generate training data, obviating the expensive process of human labeling.
\item We compile an extensive benchmark for this task by collecting 244 real test cases from user spreadsheets and online forums.\footnote{Available at \url{https://github.com/LiPengCS/Auto-Tables-Benchmark}.}  Our evaluation suggests that \at can successfully synthesize transformations for over 70\% of test cases at interactive speeds (with sub-second latency).
\end{itemize}

%% file: Related.tex
\vspace{-2mm}
\section{Related work}
\label{sec:related}

\textbf{By-example transformation using program synthesis.} There is a large body of prior work on using input/output examples to synthesize transformations. One class of techniques focuses on the so-called ``row-to-row'' transformations where one input row maps to one output row (e.g., TDE~\cite{tde} and FlashFill~\cite{flashfill}), which are orthogonal to the table-restructuring transformations in \at, because these systems do not consider operators shown in Table~\ref{tab:dsl} that can change the structure of tables. Other forms of row-to-row transformations using partial specifications (e.g., transform-by-pattern~\cite{yang2021auto, trifacta-transform-by-pattern}, transform-by-target~\cite{jin2020auto, koehler2019incorporating}, and transform-for-joins~\cite{zhu2017auto, nobari2022efficiently}), are similarly also orthogonal to the problem we study in this work.

A second class of by-example transformation consider ``table-to-table'' operators, such as Foofah~\cite{jin2017foofah} and SQL-by-example techniques like PATSQL~\cite{takenouchi2020patsql}, QBO~\cite{qbo}, and Scythe~\cite{sql-by-example}. These techniques consider a subset of table-restructuring operators, %(e.g., SQL-by-example techniques focus on SPJ queries and not restructuring steps), 
which fall short in the \at task as we will show experimentally. It is also worth pointing out that unlike \at that takes no examples, these  systems require users to provide \textit{one example output table}, which is a significant amount of effort for users.

\textbf{Computer vision models for object detection.} Substantial progress has been made in the computer vision literature on object detection, with variants of CNN architectures being developed to extract salient visual features from pictures~\cite{vgg, resnet, alexnet}. 

Given the ``visual'' nature of our problem shown in Figure~\ref{fig:combined-ex}, and the strong parallel between ``pixles'' in images and ``rows/columns'' in tables, both of which form two-dimensional rectangles,  our model architecture is inspired by CNN-architectures for object detection, but specifically designed for our table transformation task. %(e.g., the use of 2x2 filters that are suited for homogeneity test between neighboring rows/columns). 

% removed in revision
%\revised{}
%\begin{comment}
\textbf{Representing tables using deep models.} Different techniques have been proposed to represent tables using deep models  (e.g., TaBERT~\cite{yin2020tabert}, Tapas~\cite{herzig2020tapas}, Turl~\cite{deng2022turl}, etc.). Most of these focus on natural-language (NL) aspects of tables, and tailor to NL-related tasks (e.g., NL-to-SQL and entity-linking~\cite{herzig2020tapas, yin2020tabert}), which we show are not suited for our table-transformation task, as it needs to exploit the structural homogeneity of tables (e.g., cell similarity in row/column-directions.). %There is also a line of work on representing spreadsheets using rich visual features (fonts, colors, etc.)~\cite{dong2019tablesense, wang2021tuta}, which are not considered in this work as \at aims to handle general-purpose tables that may not have such spreadsheet features. 

\textbf{Database schema design.} There is a body of classical database research on schema design, which typically involves normalizing or decomposing one large table into multiple smaller tables, so that the decomposed tables satisfy relational ``normal forms'' (3NF, BCNF, etc.)~\cite{normal-forms}, that can improve storage efficiency and avoid update anomalies, among other things.  In contrast, our work has the goal of restructuring an input table to make it easy to query, which is always \textit{single-table to single-table}, and thus both orthogonal and complementary to schema design (e.g., our transformed table can then be subject to schema-design steps if it needs to be stored in databases). 
%\end{comment}

%% file: Preliminary.tex
\vspace{-2mm}
\section{Preliminary and Problem}
\label{sec:preliminary}

%\begin{comment}
\begin{figure*}[t]
\vspace{-18mm}
    \centering
    \includegraphics[width=2.0\columnwidth]{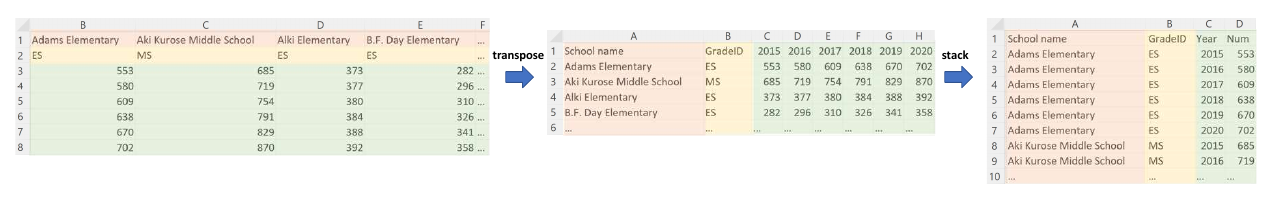}
    \vspace{-4mm}
    \caption{An example input table (on the left) that requires two transformation steps to relationalize: (1) a ``\code{transpose}'' step to swap rows and columns, (2) a ``\code{stack}'' step to collapse  homogeneous columns (C to H) into two new columns. The resulting output table (on the right) becomes substantially easier to query with SQL (e.g., to filter and aggregate).}
    \label{fig:multi-step-ex}
\vspace{-5mm}
\end{figure*}
%\end{comment}

In this section, we will introduce the table-restructuring operators considered in this work, and  describe our synthesis problem.
\vspace{-2mm}
\subsection{Table-restructuring operators}
We consider 8 table-restructuring operators in our DSL, which are listed in Table~\ref{tab:dsl}. Based on our analysis of tables in the wild (in user spreadsheets and on the web), these operators cover a majority of scenarios required to relationalize tables. Note that since our synthesis framework uses self-supervision for training that is not tied to the specific choices of operators, our approach can be easily extended to include additional operators for new functionalities.
%\kr{can you add more explanation on why these operators were selected? it is easy to extend your method to other operators that have an inverse right? }

In this section, we  will introduce the first 4 operators and their parameters shown in Table~\ref{tab:dsl} (we will give additional details in our technical report~\cite{full} in the interest of space).

% \iftoggle{fullversion}
% {
%     We will give details of the remaining operators in Appendix~\ref{apx:additional-op}, which are all similar in spirit.
% }
% {
%     We will leave an exact specification of the remaining operators to our technical report~\cite{full} in the interest of space, since they are all similar in spirit.
% }

\textbf{Stack.}  \code{Stack} is a Pandas operator~\cite{op-stack} (also known as \code{melt} and \code{unpivot} in other contexts), that collapses contiguous blocks of homogeneous columns into two new columns. Like shown in Figure~\ref{fig:combined-ex}(a), column headers of the homogeneous columns  (``\code{19-Oct}'', ``\code{20-Oct}'', etc.) are converted into values of a new column called ``\code{Date}'', making it substantially easier to query (e.g., to filter using a range-predicate on the ``\code{Date}'' column).

\underline{Parameters.} In order to properly invoke \code{stack}, one needs to provide two important parameters, \code{start\_idx} and \code{end\_idx} (listed in the third column of Table~\ref{tab:dsl}), which specify the starting and ending column index of the homogeneous column-group that needs to be collapsed. In the case of Figure~\ref{fig:combined-ex}(a), we should use \code{start\_idx}=3 (corresponding to column D) and \code{end\_idx}=12 (column M).

Note that because in \at we aim to synthesize complete transformation steps that can execute on input tables, which requires us to predict not only the operators (e.g., \code{stack} for the table in Figure~\ref{fig:combined-ex}(a)), but also the exact parameters values correctly (e.g., slightly different parameters such as \code{start\_idx}=4  and \code{end\_idx}=12 would fail to produce the desired transformation). 

\textbf{Wide-to-long.} \code{Wide-to-long} is an operator in Pandas~\cite{op-wide-to-long}, that collapses repeating column groups into rows (similar functionality can also be found in R~\cite{op-R-wide-to-long}). Figure~\ref{fig:combined-ex}(b) shows such an example, where ``\code{Revenue/Units Sold/Margin}'' from different years form  column-groups that repeat once every 3 columns. All these repeating column-groups can collapse into 3 columns, with an additional ``\code{Year}'' column for year info from the original column headers, as shown on the right in Figure~\ref{fig:combined-ex}(b). Observe that \code{wide-to-long} is similar in spirit to \code{stack} as both collapse homogeneous columns, although \code{stack} cannot produce the desired outcome when columns are repeating in groups, as is the case in this example.
%-- when all columns to collapse repeat once every 1 column, \code{wide-to-long} degenerates into the \code{stack} operator.

\underline{Parameters.} \code{wide-to-long} has 3 parameters, where \code{start\_idx} and \code{end\_idx} are similar to the ones used in \code{stack}. 
It has an additional parameter called ``\code{delim}'', which is the delimitor used to split the original column headers, to produce new column headers and data-values. For example, in the case of Figure~\ref{fig:combined-ex}(b),  ``\code{delim}'' should be specified as ``\code{ - }'' to produce: (1) a first part corresponding to values for the new ``\code{Year}'' column (``\code{2018}'', ``\code{2019}'', etc.); and (2) a second part corresponding to the new column headers in the transformed table (``\code{Revenue}'', ``\code{Units Sold}'', etc.). Like in \code{stack}, all 3 parameters here need to be instantiated correctly, before we can synthesize the desired transformation.

\textbf{Transpose.} \code{Transpose} is a table-restructuring operator that converts rows to columns and columns to rows, which is also used in other contexts such as in matrix computation.  Figure~\ref{fig:combined-ex}(c) shows an example input table on the left, for which \code{transpose} is needed to produce the output table shown on the right, which would become relational and easy to query.

\underline{Parameters.} Invoking \code{transpose} requires no parameters, as all rows and columns will be transposed.

\textbf{Pivot.} Like \code{transpose}, \code{pivot} also converts rows to columns, as the example in Figure~\ref{fig:combined-ex}(d) shows. However, in this case rows show repeating-groups (whereas in \code{wide-to-long} columns show repeating-groups), which need to be transformed into columns, like shown on the right of Figure~\ref{fig:combined-ex}(d). 

\underline{Parameters.} 
\code{Pivot} has one parameter, ``\code{repeat\_frequency}'', which specifies the frequency at which the rows repeat in the input table. In the case of Figure~\ref{fig:combined-ex}(d), this parameter should be set to 4, as the color pattern of rows would suggest.

% removed in revision
%\revised{}
%\begin{comment}
\textbf{Additional operators.}
Table~\ref{tab:dsl} has 4 additional table-restructuring operators, which we will briefly mention here. These include (1): ``\code{\textbf{explode}}''~\cite{op-explode}, which converts columns with composite values (violating the First Normal Form~\cite{codd1990relational}) into atomic values, so that the table can be queried using standard SQL; (2): ``\code{\textbf{ffill}}''~\cite{op-ffill} that fills values in structurally empty cells so that the table can be queried;  (3): ``\code{\textbf{subtitle}}'' that converts  rows representing table sub-titles into separate columns for ease of queries; and finally (4): a ``\code{\textbf{none}}'' operator for input tables that are already relational, for which no transformation is needed, which is needed explicitly so that we do not ``over-trigger'' on tables that require no transformation. 
% \iftoggle{fullversion}
% {
%     We leave details of these operators to Appendix~\ref{apx:additional-op}, as they are similar in nature to the operators we already described.
% }
% {
%     We leave details of these operators to a technical report~\cite{full} in the interest of space, as they are similar in nature to the operators we already described.
% }
% %\end{comment}

\begin{comment}
\stitle{DSL.} We consider the following 8 operators in our experiments. 
\begin{table}[!h]
\scalebox{0.6}{
\begin{tabular}{l|l|l}
\toprule
Operator & Pandas API & \at{} DSL \\
\midrule
Transpose & df.T & dsl\_transpose(df) \\
Stack & pd. melt(df, id\_vars, value\_vars) & dsl\_stack(df, start\_idx, end\_idx) \\
Wide to long & pd.wide\_to\_long(df, stubnames, i) & dsl\_wide\_to\_long(df, start\_idx, end\_idx) \\
Explode & df.explode(column) & dsl\_wide\_to\_long(df, column\_idx) \\
Ffill & df{[}column{]}.ffill() & dsl\_wide\_to\_long(df, end\_idx) \\
Pivot & df.pivot(index, columns, values) & dsl\_pivot(df, row\_frequency) \\
Subtitles & - & dsl\_subtitle(df) \\
None & df\_out = df & dsl\_none(df) \\
\bottomrule
\end{tabular}
}
\end{table}
\end{comment}

\subsection{Problem statement}

Given these table-restructuring operators  listed in Table~\ref{tab:dsl}, we now introduce our synthesis problem as follows.
\vspace{-1mm}
\begin{definition}
\label{def:problem}
%[\at] 
Given an input table $T$, and a set of operators $\mathbf{O} = \{stack, transpose, pivot, \ldots \}$, where each operator $O \in \mathbf{O}$ has a parameter space $P(O)$. Synthesize a sequence of multi-step transformations $M = (O_1(p_1), O_2(p_2), \ldots, O_k(p_k))$, with  $O_i \in \mathbf{O}$ and $p_i \in {P(O_i)}$ for all $i \in [k]$, such that applying each step $O_i(p_i) \in M$  successively on $T$ produces a relationalized version of $T$.
\end{definition}

Note that in our task, we need to predict both the operator $O_i$ and its exact parameters $p_i$ correctly, each step along the way. This is challenging as the search space is large --
even for a single-step transformation, there are thousands of possible operators/parameters to choose from (e.g., a table with 50 columns that requires ``\code{stack}'' will have 50x50 = 2500 possible parameters of start\_idx and end\_idx); for two-step transformations the search space is already in the millions (e.g., for ``\code{stack}'' alone it is $2500^2 \approx 6M$). Given the large search space, even
a small difference in parameters can render the resulting transformation incorrect, as shown below.

%\yeye{mention almost unique here}
%\kr{did you discuss/introduce the no-op option?}
\vspace{-1mm}
\begin{example} 
\label{ex:multi-step}
Given the input table $T$ shown on the left of Figure~\ref{fig:multi-step-ex}, the ground-truth transformation $M$ to relationalize $T$ has two-steps: $M = (\text{transpose()},$ $\text{stack(start\_idx:``2015'',}$ 
 $\text{end\_idx:``2020''}) )$. Here the first step ``\code{transpose}'' swaps the rows with columns, and the second step ``\code{stack}''  collapses the homogeneous columns (between column ``\code{2015}'' and ``\code{2020}''). Note that this is the only correct sequence of steps -- reordering the two steps, or using slightly different parameters (e.g., start\_idx=``2016'' instead of ``2015''), will all lead to incorrect output, which makes the  problem challenging. 
 
 Also note that although we show synthesized programs using our DSL syntax, the resulting programs can be easily translated into different target languages, such as Python Pandas or R, which can then be directly invoked.
 We should also note that two syntactically different programs $M_1$ and $M_2$ may be semantically equivalent, which can be verified under a set of algebraic rules.  \footnote{For example, pivot is equivalent to transpose followed by wide-to-long, and wide-to-long is equivalent to stack-split-pivot. Furthermore, the order of ffill and stack/wide-to-long can be swapped, as long as they operate on disjoint subsets of columns, etc. In our synthesis, we consider synthesized programs that are semantically equivalent to the ground-truth program also correct.}

 %\yeye{see if we need to add actual synthesized code} 
 
 %\yeye{Peng: please remind me why we call this Stack and not Melt again? The API \url{https://pandas.pydata.org/docs/reference/api/pandas.DataFrame.stack.html} does not seem to have start-idx, end-idx (unlike Melt). If we need to translate this into Pandas, do we need to first promote the column-headers between start-idx and end-idx, into column-index in Pandas, and then apply Stack? (I was thinking about writing out the translated Python code, but found this to be difficult...)}
\end{example}

%% file: Method.tex
\section{\at: Learn-to-synthesize}

We now describe our proposed \at system, which learns to synthesize transformations. We will start with an architecture overview before we delve into individual components.

\subsection{Architecture overview}

\begin{figure}[t!]
    \vspace{-3mm}
\centering\includegraphics[width=1.0\columnwidth]{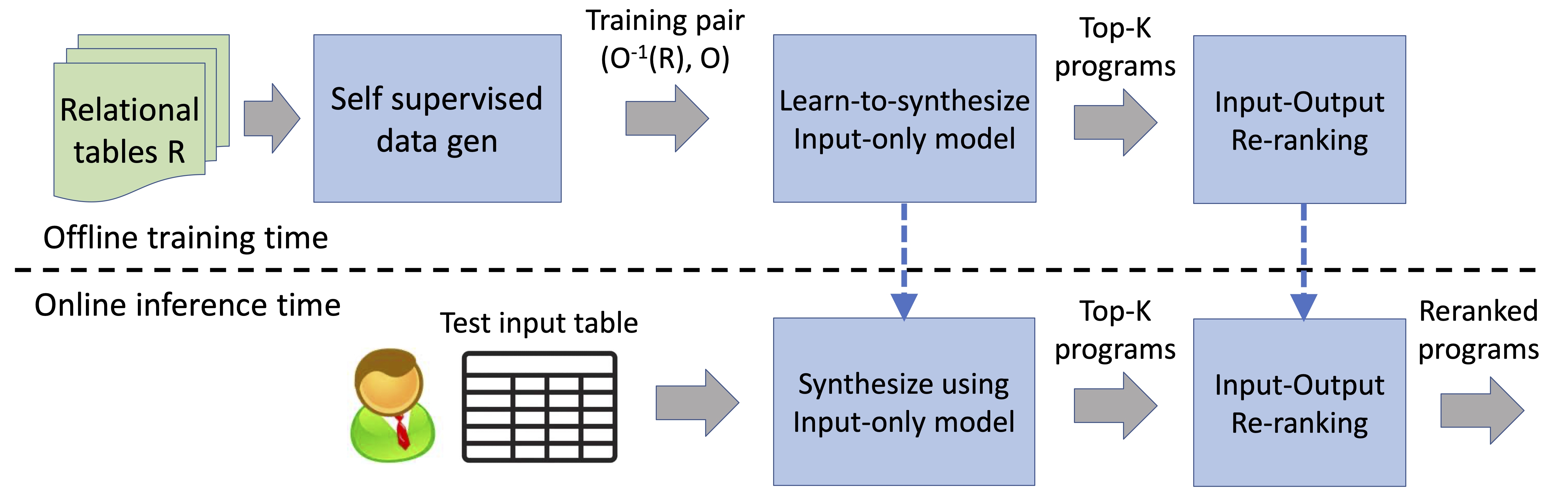}
    \vspace{-8mm}
    \caption{Architecture overview of \at}
\label{fig:architecture}
\vspace{-6mm}
\end{figure}

We represent our overall  architecture in
Figure~\ref{fig:architecture}. The system operates in two modes, with  the upper-half of the figure showing the offline training-time pipeline, and the lower-half showing the online inference-time steps.

At offline training time, \at uses three main components: (1) A ``training data generation'' component that consumes large collections of relational tables $R$, to produce (example, label) pairs; (2) An ``input-only synthesis'' module that learns-to-synthesize using the training data, and (3) An ``input-output re-ranking'' module that holistically considers both the input table and the output table (produced from the synthesized program), to find the most likely program. 

The online inference-time part closely follows the offline steps, where we directly invoke the two models trained offline (the last two blue boxes shown in the figure). When given an input table from users, we pass the table through our input-only synthesis model, to identify top-$k$ candidate programs, which are then re-ranked by the input-output model for final predictions.

We now describe these three modules in turn below.

\subsection{Self-supervised training data generation}
\label{sec:training_data_generation}

As discussed earlier, the examples in Figure~\ref{fig:combined-ex} demonstrate that there are clear patterns in the input tables that we can exploit (e.g., repeating column-groups and row-groups) to predict required transformations for a given table. Note that these patterns are ``visual'' in nature, which can likely be captured by computer-vision-like algorithms.\footnote{Like computer vision problems such as object detection where hand-crafted heuristics are hard to write, the row/column-level patterns existing in our tables are also hard to write with heuristics, which makes a learning-based method necessary.}

The challenge however, is that unlike computer vision tasks that typically have large amounts of training data (e.g., ImageNet~\cite{imagenet}) in the form of (image, label) pairs, in our synthesis task, there is no existing labeled data that we can leverage. Labeling  tables manually from scratch are likely too expensive to scale.

\underline{Leverage inverse operators.} To overcome the lack of data, we propose a novel self-supervision framework leveraging the inverse functional-relationships between operators, to automatically generate large amounts of training data without using humans labels.

\begin{figure}[t!]
\vspace{-9mm}
    \centering
    \includegraphics[width=1.0\columnwidth]{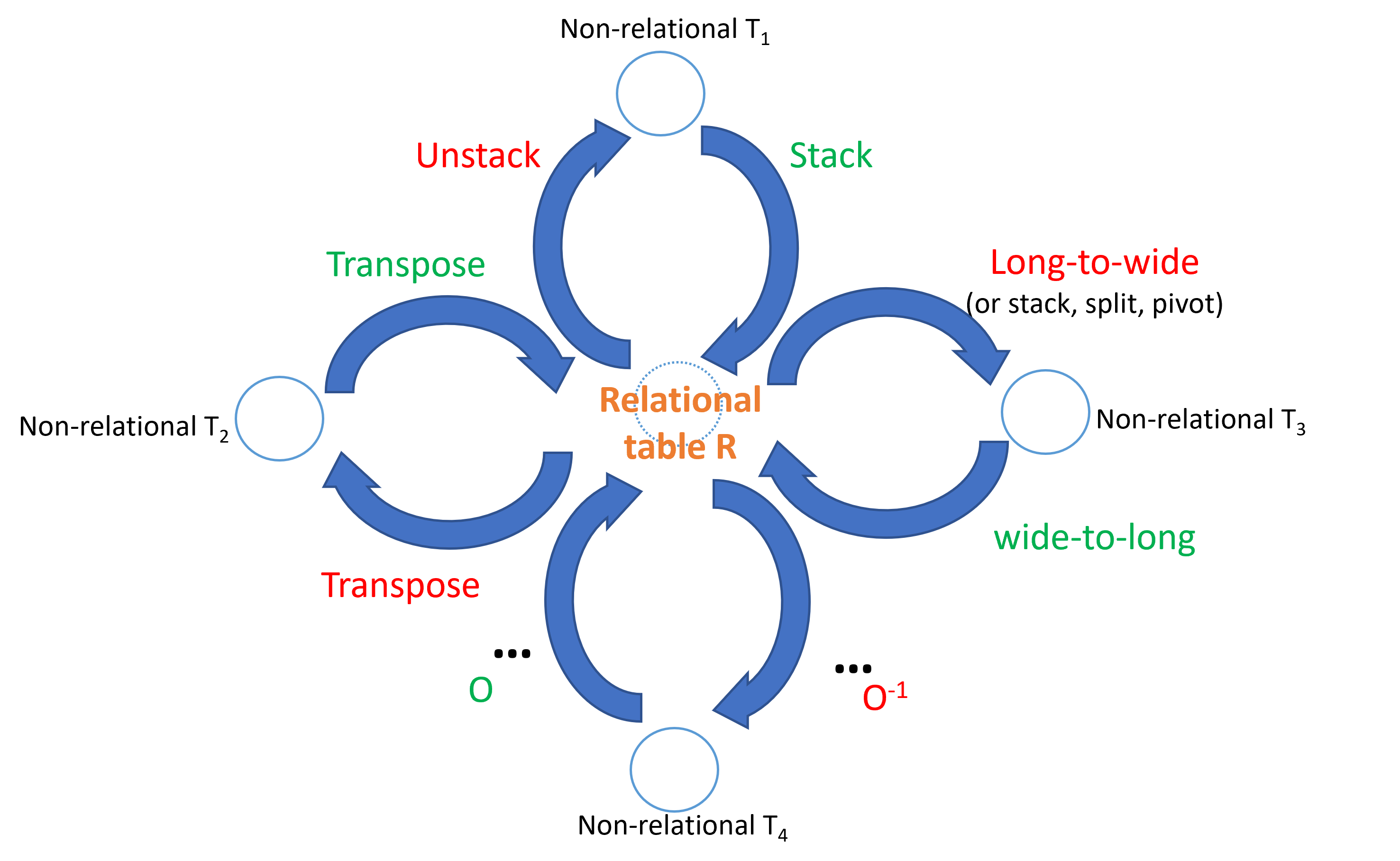}
    \vspace{-8mm}
    \caption{Leverage inverse operators to generate training data. In order to learn-to-synthesize operator $O$, we can start from any relational table $R$, apply its inverse operator $O^{-1}$ to obtain $O^{-1}(R)$.  Given $T = O^{-1}(R)$ as an input table, we know $O$ must be its ground-truth transformation, because $O(O^{-1}(R)) = R$.}
    \label{fig:inverse-function}
\vspace{-5mm}
\end{figure}

Figure~\ref{fig:inverse-function} shows the overall idea of this approach. For each operator $O$ in our DSL that we want to learn-to-synthesize, we can find its inverse operator (or construct a sequence of steps that are functionally equivalent to its inverse), denoted by $O^{-1}$. For example, in the figure we can see that the inverse of ``\code{transpose}'' is ``\code{transpose}'', the inverse of ``\code{stack}'' is ``\code{unstack}'', while the inverse of ``\code{wide-to-long}'' can be constructed as a sequence of 3 steps (``\code{stack}'' followed by ``\code{split}'' followed by ``\code{pivot}'').

The significance of the inverse operators, is that it allows us to automatically generate training examples. 
Specifically, to build a training example for operator $O$ (e.g., ``\code{stack}''), we can sample any relational table $R$, and  apply the inverse of $O$, or $O^{-1}$ (e.g., ``\code{unstack}''), to generate a non-relational table $T = O^{-1}(R)$. For our task, given $T$ as input, we know $O$ must be its ground-truth transformation, since by definition $O(T) = O(O^{-1}(R))=R$, and $R$ is known to be relational. 
This thus allows us to generate $(T, O)$ as an (example, label) pair, which can be used for training. %, with $T$ as the input table and $O$ the ground-truth transformation.

Furthermore, we can easily produce such training examples at scale, by sampling: (1) different relational tables $R$; (2) different operators $O$; and (3) different parameters associated with each $O$, therefore addressing our lack of data problem in \at.

The overall steps of the data generation process are shown in Algorithm~\ref{alg:data-gen}, where Line~\ref{line:sample_op}, Line~\ref{line:sample_table}, Line~\ref{line:sample_parameter} correspond to the sampling of operators ($O$), tables ($R$), and parameters ($p$), respectively, that together creates diverse training examples.  We note that in Line~\ref{line:augment_table}, we perform an additional ``data augmentation'' step to create even more diversity in training, which we explain below.
%\kr{So the data generation only considers one step of the transformation?}

\underline{Data Augmentation.} Data augmentation~\cite{shorten2019survey} is a popular technique in computer vision and related fields, to enhance training data and improve model robustness. 
%used for image data or text data to enlarge the size of training examples and improve the robustness of deep learning models. 
For example, in computer vision tasks, it is observed that training using additional data generated from randomly flipped/rotated/cropped images, can lead to improved model performance (because an image that contains an object, say dog, should still contain the same object after it is flipped/rotated, etc.)~\cite{shorten2019survey}.

In the same spirit, we
augment each of our relational table $R$ by (1) Cropping, or randomly sampling contiguous blocks of rows and columns in $R$ to produce a new table $R'$;  and (2) Shuffling, or randomly reordering the rows/columns in $R$ to create a new $R'$. In \at, we start from over 15K relational tables crawled from public sources (Section~\ref{sec:exp}), and create around 20 augmented tables for each relational table $R$. This further improves the diversity of our training data and end-to-end model performance, as we will show in the experiments. %(Section~\ref{subsec:ablation}). 

%\yeye{see if we need to add an example to show how parameters are selected. maybe not, discussing how to select parameters will necessitate the explanation of heuristics.}

%First, since any subset of a relational table is still relational, we can randomly sample contiguous blocks of rows and columns from the original table to generate new relational tables. Second, since the order of rows or columns in a relational table is insignificant, we can randomly changing the order of columns and rows to get new tables. Third, for operators with parameters, we can generate multiple training examples from one relation table by picking different parameters for the inverse operator. For example, for unstack, its parameter specifies the column that will be expanded. By choosing different columns, we can generate different training tables. As we will show in our experiments, we generate 10$\sim$20 tables from each relational table for each operator and the data augmentation significantly improves the model performance.

\begin{small}
\begin{algorithm}[t]
\SetKw{kwReturn}{return}
 \Input{DSL operators $\mathbf{O}$, large collections of relational tables $\mathbf{R}$}
 \Output{Training table-label pairs: $(T, O_p)$}
 
 %$V \leftarrow \{v_T | T \in \mathbf{T} \}$, with $v_T$ representing each $T \in \mathbf{T}$ %\tcp{Algorithm~\ref{algo:constraints}} 

 $E \leftarrow \{\}$
 
  \ForEach{$O$ in $\mathbf{O}$ \label{line:sample_op}} 
    {
      \ForEach{$R$ in $\mathbf{R}$ \label{line:sample_table}} 
        {
            \ForEach{$R'$ in \text{Augment}($R$) \label{line:augment_table} \tcp{Crop rows and columns} } 
            {
               $p \leftarrow $ sample valid parameter from space $P(O)$ \label{line:sample_parameter} 
               
               $O^{-1}_{p'} \leftarrow $ construct the inverse of $O_p$ \label{line:construct_inverse}
               
               $T \leftarrow O^{-1}_{p'}(R')$

               $E \leftarrow E \cup \{(T, O_p)\}$
           }
       }
    %\tcp{set edge-weight}
    }

\kwReturn all training examples $E$
\caption{Auto-gen training examples}
\label{alg:data-gen}
\end{algorithm}
\end{small}

\begin{figure*}[t]
    \centering
    \vspace{-15mm}
    \includegraphics[width=2\columnwidth]{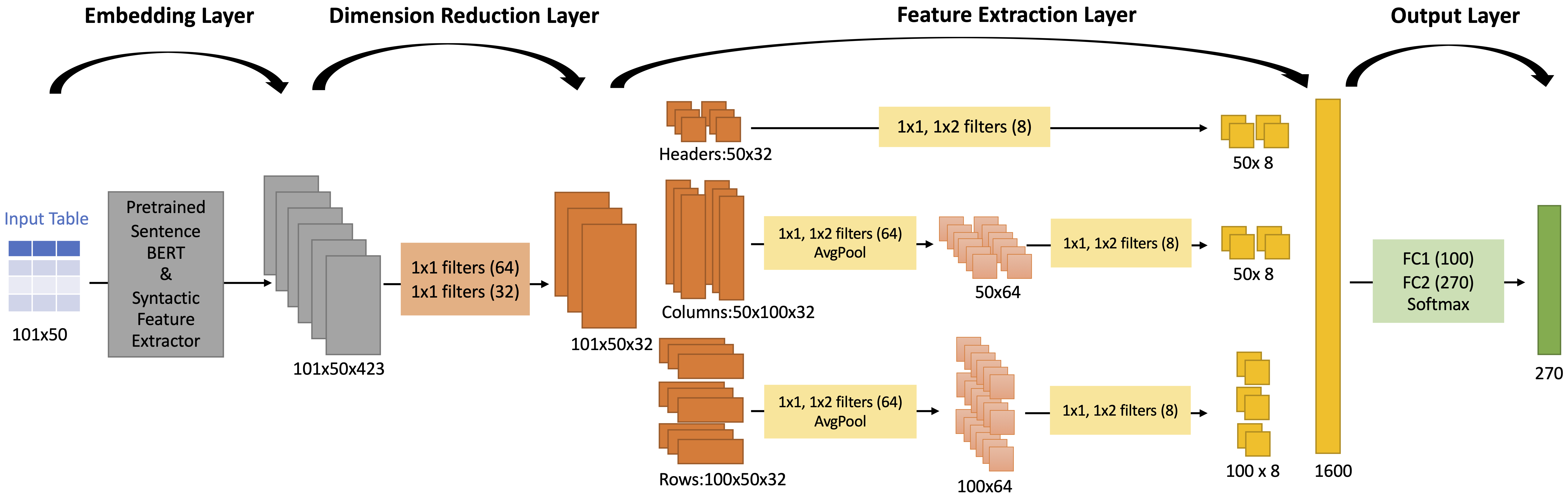}
    \vspace{-4mm}
    \caption{Input-only synthesis: model architecture. %We use a fixed ``window'' with the first 100 data-rows (plus a header) and 50 columns at the top-left corner of each input $T$, like in computer-vision.
    }
    \label{fig:input_model_arch}
    \vspace{-3mm}
\end{figure*}

\vspace{-2mm}
\subsection{Input-only Synthesis}
\label{subsec:input-model}
%\yeye{may need to see how to add some intro of cnn}

After obtaining large amounts of training data in the form of $(T, O_p)$ using self-supervision, we now describe our ``input-only'' model that takes $T$ as input, to predict a suitable transformation $O_p$.

\subsubsection{\textbf{Model architecture}} \hfill\\ 
We develop a computer-vision inspired model specifically designed for our task, which scans through rows and columns to extract salient tabular features, reminiscent of how computer-vision models extract features from image pixels for object detection.

Our model architecture in Figure~\ref{fig:input_model_arch} consists of four sets of layers: (1) table embedding, (2) dimension reduction, (3) feature extraction, and (4) output layers. We will describe each in turn below.

\underline{Table embedding layers.} Given an input table $T$, the embedding layer encodes each cell in $T$ into a vector, to obtain an initial representation of $T$ for training. At a high level, for each cell we want to capture both (1) the ``\textit{semantic features}'' (e.g., people-names vs. company-names), and (2) the ``\textit{syntactic feature}'' (e.g., data-type, string-length, punctuation, etc.), because both semantic and syntactic features provide valuable signals in our task, e.g., in determining whether rows/columns are homogeneous or similar. 

For semantic features, we use the pre-trained Sentence-BERT~\cite{reimers-2019-sentence-bert} (a state-of-the-art embedding in NLP), which maps each cell into a 384-dimension vector that encodes its semantic meaning. For syntactic features, we encode each cell using 39 pre-defined syntactic attributes (data types, string lengths, punctuation, etc.). Concatenating the syntactic and semantic features produces a 423-dimension vector for each cell. For an input table $T$ with $n$ rows and $m$ columns, this produces a $n \times m \times 423$ tensor as its initial representation.\footnote{Like in computer vision problems that use a fixed ``window'', we take the first 100 data-rows (plus a header) and 50 columns at the top-left of each input $T$ (producing a $101 \times 50 \times 423$ tensor), which is sufficient to identify table patterns and predict transformations, like the examples in Figure~\ref{fig:combined-ex} would show.} % (For smaller tables, we pad empty cells with all-0 vectors.)} 

\begin{figure}
    \centering
    \begin{subfigure}[b]{\columnwidth}
        \centering
        \includegraphics[width=\columnwidth]{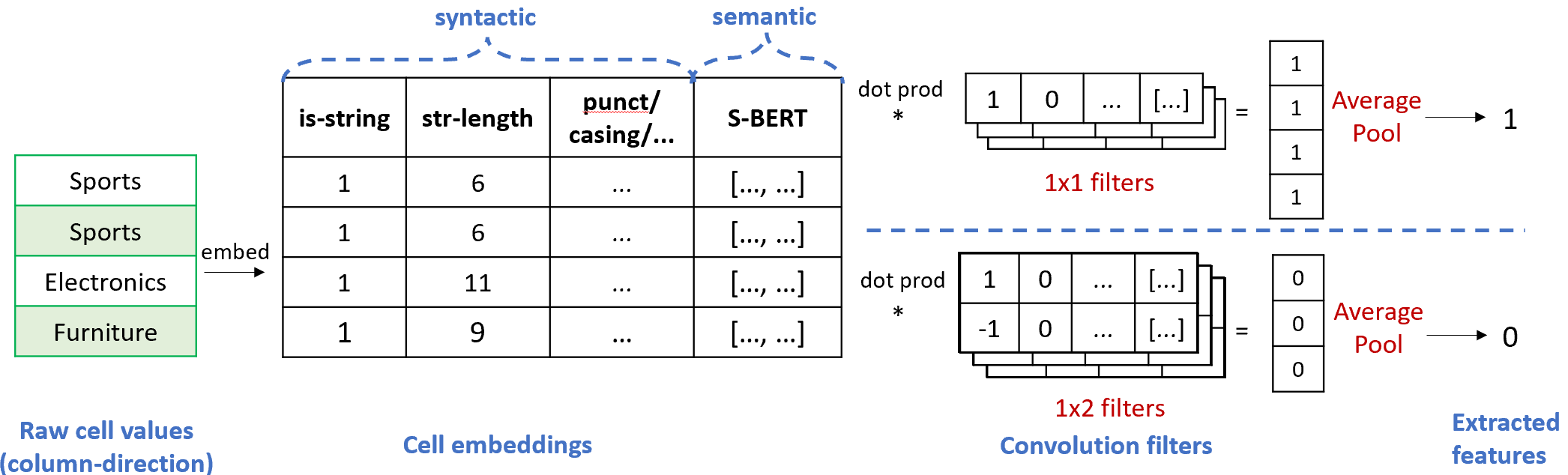}
        \caption{Feature extraction for the input table in Figure~\ref{fig:combined-ex}(a), Column-B}
    \end{subfigure}
        \begin{subfigure}[b]{\columnwidth}
        \centering
        \includegraphics[width=\columnwidth]{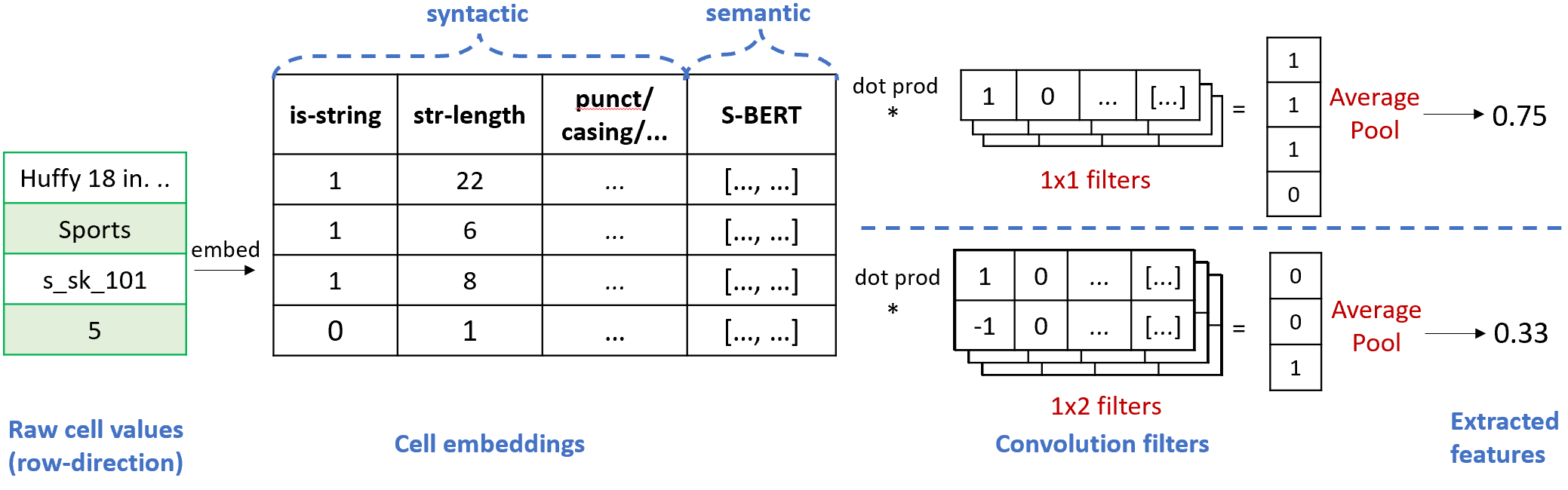}
        \caption{Feature extraction for the input table in Figure~\ref{fig:combined-ex}(a), Row-2}
    \end{subfigure}
    % \begin{subfigure}[b]{0.9\columnwidth}
    %     \centering
    %     \includegraphics[width=\columnwidth]{figures/feature_extraction_ex2.png}
    %     \caption{Extracted high-level features}
    % \end{subfigure}
    \vspace{-7mm}
    \caption{Example feature extraction %on input table in Figure~\ref{fig:combined-ex}(a), 
    using 1x1 and 1x2 filters}
    \vspace{-5mm}
    \label{fig:example_feature_extraction}
\end{figure}

The left half of Figure~\ref{fig:example_feature_extraction} shows a simple sketch of this embedding step, which we will explain in more detail later.

\underline{Dimension reduction layers.} Since the initial representation from the pre-trained Sentence-BERT has a large number of dimensions (with information likely not needed for our task, which can slow down training and increase the risk of over-fitting), 
we add dimension-reduction layers  using two convolution layers with $1\times 1$ kernels, to reduce the dimensionality from  423 to 64 and then to 32, to produce $n \times m \times 32$ tensors. Note that we explicitly use $1\times 1$ kernels so that the trained weights are shared across all table-cells, to produce consistent representations after dimension reduction.

%the initial representation, while preserving salient information for our task.  For this purpose, we use two sequential 1D convolution layers with kernal size $1\times 1$, to reduce the dimension of each embedding vector from 423 to 64 and then to 32. The output of this layer will be a $m\times n \times 32$ tensor.

\underline{Feature extraction layers.} We next have feature extraction layers that are reminiscent of CNN~\cite{li2021survey} but specifically design for our table task. Recall from Figure~\ref{fig:combined-ex} that the key signals for our task are: 
\begin{itemize}[leftmargin=*,noitemsep,topsep=0pt]
\item (1) identify whether values in row or column-directions are ``similar'' enough to be ``homogeneous'' (e.g., Figure~\ref{fig:combined-ex}(b) vs. Figure~\ref{fig:combined-ex}(c)); 
\item (2) identify whether entire rows or columns are ``similar'' enough to show repeating patterns (e.g., Figure~\ref{fig:combined-ex}(b) vs. Figure~\ref{fig:combined-ex}(d)).
\end{itemize}

Intuitively, if we were to hand-write heuristics, then signal (1) above can be extracted by comparing the representations of adjacent cells in row- and column-directions. 
On the other hand, signal (2)  can be extracted by computing the  average representations of each row and column, which can then be used to find repeating patterns. 

%he feature extraction layer aims to extract useful and informative features from table embeddings. Our intuition is that the table relationlization task highly relies on the structural information within and across column/rows. For example, if there are a great number of delimiters (e.g., commas) within a column, it is a strong signal to perform an explode operator; if a column contains diverse values while the rows contains homogeneous values, it is a strong signal to perform a transpose operator; if values across multiple columns are homogeneous , it is a strong signal to perform a stack operator. 

Based on this intuition, and given the strong parallel between the row/columns in tables and pixels in images, we design feature-extraction layers inspired by \textit{convolution filters}~\cite{li2021survey} that are popular in CNN architectures to extract visual features from images~\cite{vgg, alexnet}. 
Specifically, as shown in Figure~\ref{fig:input_model_arch}, we use 1x2 and 1x1 convolution filters followed by average-pooling, in both row- and column-directions, to represent rows/columns/header. Unlike general $n$x$m$ filter used for image tasks (e.g., 3x3 and 5x5 filters in VGG~\cite{vgg} and ResNet~\cite{resnet}), our design of filters are tailored to our table task, because:
\begin{itemize}[leftmargin=*,noitemsep,topsep=0pt]
\item (a) 1x2 filters can easily learn-to-compute signal (1) above (e.g., 1x2 filters with +1/-1 weights can identify the representation differences between  neighboring cells, which when averaged, can identify homogeneity in row/column directions).  
\item (b) 1x1 filters can easily learn-to-compute signal (2) above (e.g., 1x1 filters with +1 weights followed by average-pooling, correspond to representations for entire rows/columns, which can be used to find repeating patterns in subsequent layers).
\end{itemize}

%As shown in Figure~\ref{fig:input_model_arch},  using the 1x1 and 1x2 filters, we represent columns, rows and the headers separately, through two layers of 64 and 8 filters with trainable filter weights, where different filters can learn and balance the relative importance of different cell-level features.

%In the end, we produce 50x8 representation for 50 input columns (encoding column homogeneity and column content), and similarly 100x8 representation for 100 input rows (for row homogeneity and row content), which are fed into the final output layers of our model.

We use an example below to demonstrate why these 1x1 and 1x2 filters are effective for extracting tabular features.
\vspace{-1mm}
\begin{example}
\label{ex:filters}
%We demonstrate the use of these filters for extracting features in tables, using an example shown in Figure~\ref{fig:example_feature_extraction}. %, based on the input table shown in Figure~\ref{fig:combined-ex}(a). 
%Recall that our CNN-inspired architecture uses convolution filters to scan line-by-line, in both row and column directions. %(illustrated in the lower-half of the feature-extraction layers in Figure~\ref{fig:input_model_arch}).
Figure~\ref{fig:example_feature_extraction}(a) shows a simplified example, when using Column-B of Figure~\ref{fig:combined-ex}(a) as input, which has a list of values ``\code{Sports}'', ``\code{Electronics}'', etc. These raw cell values first pass through the embedding step, which produces a row of features for each value, with both \textit{syntactic features} (under the headers ``\code{is-string}'', ``\code{str-length}'', etc.), and \textit{semantic features} (under the header ``\code{s-BERT}'' for sentence-BERT). This results in an embedding table, where each row corresponds to an input cell.

Next, we pass this embedding table  through 1x1 and 1x2 convolution filters, which performs element-wise dot-product~\cite{li2021survey}. Assuming we have a simple 1x1 filter shown at the top of the figure, with weights $[1, 0, \ldots]$.  Because only the first bit of this simple filter is $1$ and the rest is $0$, performing a dot-product on the embedding table essentially only extracts the ``\code{is-string}'' type information of each cell, which in this case is all $1$, leading to a matrix of $[1, 1, 1, 1]$ (since all cells are of type string). After average pooling, this results in a single feature-value $1$ to represent a specific aspect of this entire column (in this case, type information).

We should note that this is just one example 1x1 filter -- there exists many such 1x1 filters (shown as stacked in the figure), all of which have learned weights that extract different aspects of syntactic/semantic information from input cells (string-length, semantic-meaning, etc.), thus forming a holistic representation of values in the column, to facilitate downstream comparison of ``similar'' columns (e.g., to identify repeating rows/columns), as mentioned above as signal (2) for our task.

The 1x2 filters, on the other hand, work to ``compare'' adjacent values in the same column, which intuitively test for homogeneity. For instance, assuming there is a simple 1x2 filter with only +1 and -1 weights in the first column, as shown in the figure. Performing a dot-product in this case ``compares'' the ``\code{is-string}'' type info for neighboring cells, using a sliding window for rows from the top to bottom, which results in $[0, 0, 0]$ (because the convolution computes $1*1 + 1*(-1) = 0$). This is again averaged to produce a feature-value $0$, indicating no type difference, and thus good homogeneity, in the list of given values in the column-direction.

This is again only one example 1x2 filter -- there are many other 1x2 filters with different learned-weights (stacked in the figure) that use different syntactic/semantic features to test for homogeneity between neighboring cells, which corresponds to the signal (1) we want to extract as mentioned earlier.

Recall that our CNN-inspired architecture uses convolution filters to scan line-by-line, in both row and column directions. So in the row-direction our filters work in a similar manner.

The same operations in row-direction is shown in Figure~\ref{fig:example_feature_extraction}(b), which uses Row-2 of the table in Figure~\ref{fig:combined-ex}(a) as example. In this case we have a list of heterogeneous cell values ``\code{Huffy 18 in.}'', ``\code{Sports}'', ``\code{s\_sk\_101}'', ``\code{5}'', etc. In this case, performing a dot-product using the same 1x2 filter produces a feature-vector of $[0, 0, 1]$ (note that the last entry is $1$ because the   ``\code{is-string}'' value for the last two input cells are 1 and 0, leading to a convolution of $1*1 + (-1)*0 = 1$). Average-pooling would then produce 0.33 here, indicating inconsistent types for the list of values in the row-direction (0 would indicates homogeneity, with +1/-1 filter-weights). Other 1x2 filters would work in similar manners, to identify more signals of heterogeneity in the row-direction, all of which are important ingredients to identify latent patterns in the table and corresponding transformations. 

These first-level of features-values from row/column-directions will then go through a second-level of 1x1 and 1x2 convolution filters, to compare and identify similar rows/columns (based on row/column representation from 1x1 filters), to ultimately reveal repeating rows and columns like the color-coded patterns show in Figure~\ref{fig:combined-ex}. These tabular features will pass down to the next output layers, for final classifications.  
\end{example}

\underline{Output layers.} Our output layers use two fully connected layers followed by softmax classification, as shown in Figure~\ref{fig:input_model_arch}, which produces an output vector that encodes both the predicted operator-type, and its parameters.  For example, since we consider 8 possible operator types in our DSL, we encode this as a 8-dimension one-hot vector. 
Similarly, we represent parameters of each operator as additional bits in the same output vector, resulting in an output vector of 270 dimensions, which in effect makes multiple predictions (operator-type and parameters) simultaneously, for a given $T$.

%, e.g., the parameter ``\code{stack\_start\_idx} (ssi)'' takes the possible range of [1, 50] (since we use 50 left-most input columns to predict transformations) and thus can be encoded as a 50-dimension one-hot vector. 

%Concatenating these possible choices of operator types and parameters together yields an output vector of 270 dimensions, where different segments of the vector encode different operator and parameter predictions, like shown in  Table~\ref{tab:label_vector_encode}. Note that while in our problem we use a 270-dimension 0/1 vector to represent our space of possible choices in synthesizing transformations, the same scheme is general and can easily extend to other synthesis problems that has a different set of operators. 

We apply standard softmax functions~\cite{murphy2012machine} on each prediction vector, %that corresponds to a different prediction (operator-type or parameters),
so that the output of each prediction is normalized into a probability distribution.

% \begin{table}[t]
% \scalebox{0.7}{
% \begin{tabular}{cc|cc}
% \toprule
% \textbf{Prediction} & \textbf{Vector Position} & \textbf{Prediction} & \textbf{Vector Position} \\
% \midrule
% operator type ($O$) & 1:8 & wtl end idx (wei) & 159:208 \\
% stack start idx (ssi) & 9:58 & explode column idx (eci) & 209:258 \\
% stack end idx (sei) & 59:108 & pivot row freq (prf) & 259:268 \\
% wtl start idx (wsi) & 109:158 & ffill end idx (fei)& 269:270 \\
% \bottomrule
% \end{tabular}
% }
% \caption{Encoding of operator types and parameters, using different parts of one output vector}
% \vspace{-6mm}
% \label{tab:label_vector_encode}
% \end{table}

%\stitle{Training and inference.}

\subsubsection{\textbf{Training and inference}} \hfill \\ 
We now describe how we train this model shown in Figure~\ref{fig:input_model_arch}, and at inference time, use it to synthesize transformations. 

\stitle{Training time: Loss Function.} Given a training input table $T$, its ground truth operator $O$ and corresponding parameters $P=(p_1, p_2, ...)$, let $\hat{O}$ and $\hat{P} = (\hat{p}_1, \hat{p}_2, ...)$  be the model predicted probability distributions of $O$ and $P$ respectively. 
The training loss on $T$ can be computed as the sum of loss on all predictions (both the operator-type, and parameters relevant to this operator):
\begin{small}
\begin{align}
    Loss(T) = L(O, \hat{O}) + \sum_{p_i \in P, \hat{p}_i \in \hat{P}}L(p_i, \hat{p_i})
    \label{eqn:loss_one_example}
\end{align}
\end{small}
Here $L(y, \hat{y})$ denotes the cross-entropy loss~\cite{murphy2012machine} commonly used in classification -- let $y$ be a $n$-dimensional ground truth one-hot vector, and $\hat{y}$ a model predicted vector, $L(y, \hat{y})$ is defined as:
\begin{small}
\begin{align}
    L(y, \hat{y}) = -\sum_{i=1}^n y_i log(\hat{y}_i)
    \label{eqn:cross_entropy}
\end{align}
\end{small}
Given large amounts of training data $\mathbf{T}$ (generated from our self-supervision  in Section~\ref{sec:training_data_generation}), we train our \at model by minimizing the overall training loss $\sum_{T \in \mathbf{T}}{Loss(T)}$ using gradient descent until convergence. We will refer to this trained model as $H$.

% Note that our model will also make predictions for parameters of other operators (e.g., our model will always output predictions for stack\_start\_idx no matter the input table needs t). hese predictions are simply ignored for loss.

% We compute the loss 
% for the prediction 

% for each training example as the sum of the cross entropy loss for operator type and parameters.

% Given a training input table $T$ generated by Algorithm~\ref{alg:data-gen}, by construction, the ground truth will be a single-step pipeline denoted by $M = {(O(p))}$

% let $O$ denote the ground truth operator type and $P_{O} = \{P_{O1}, P_{O2} ...\}$ denote the ground truth parameters.

% We consider the prediction for the operator type and parameters as individual classification tasks. For each classification task, the loss between the prediction and the ground truth can be computed using cross entropy loss. Given a predicted probability distribution $\bm{p}$ and 

% \begin{align}
% \begin{split}
% Loss = & L(y_{op}, p_{op}) \\ 
% &+ [L(y_{ssi}, p_{ssi}) + L(y_{sei}, p_{sei})] \cdot \mathbb{1} (y_{op} = stack) \\
% &+ [L(y_{wsi}, p_{wsi}) + L(y_{wei}, p_{wei})] \cdot \mathbb{1} (y_{op} = wtl) \\
% &+ L(y_{ei}, p_{ei}) \cdot \mathbb{1} (y_{op} = explode) \\
% &+ L(y_{prf}, p_{prf}) \cdot \mathbb{1} (y_{op} = pivot) \\
% &+ L(y_{fei}, p_{fei}) \cdot \mathbb{1} (y_{op} = ffill) \\
% \end{split}
% \end{align}

\stitle{Inference time: Synthesizing transformations.} 
At inference time, given an input $T$, our model $H$ produces a probability for any candidate step $O_P$ that is instantiated with operator $O$ and parameters $P = (p_1, p_2, \ldots)$,
%for each operator $O$, and probability $Pr(p)$ for each parameter $p$. Then for a predicted step $O(P)$ with operator $O$ and parameters $P = (p_1, p_2, \ldots)$, we can compute the probability of this step $O(P)$ given $T$, 
denoted by $Pr(O_P | T)$, as: 
%\begin{small}
\begin{align}
    Pr(O_P | T) = Pr(O) \cdot \prod_{p_i \in P} Pr(p_i)
\label{eqn:prob_operator}
\end{align}
%\end{small}

Using the predicted probabilities, finding the most likely transformation step $O_P^*$ given $T$ is then simply:
\begin{align}
O_P^* = \argmax_{O, P}{Pr(O_P|T)}
\label{eqn:argmax_one_step}
\end{align}

%Enumerating different combinations of operator $O$ and parameter $P$ then allows us to different steps with probabilities  $Pr(O(P) | T)$, from which we produce the top-$K$ most likely ones as our predicted transformations.

This gives us the most likely one-step transformation given $T$. As we showed in Figure~\ref{fig:multi-step-ex}, certain tables may require multiple transformation steps for our task.

%Recall that from the output vector of our model, we can compute probability of each operator $Pr(O)$, and its parameter $Pr(p)$, such that for each possible step $O_P$ instantiated by operator $O$ and parameters $P$, we can compute its likelihood $Pr(O_P)$ as: 
%\begin{align}
%    Pr(O_P) = Pr(O) \cdot \prod_{p \in P} Pr(p)
%\label{eqn:prob_operator}
%\end{align}
%where $Pr(O)$ is the probability of the operator type $O$, and $Pr(p)$ is the probability of a parameter $p \in P$.

To synthesize multi-step transformations, intuitively we can invoke our predictions step-by-step  until no suitable transformation can be found. Specifically, given an input table $T$, at step (1) we can find the most likely transformation $O_P^1$ for $T$ using Equation~\eqref{eqn:argmax_one_step}, such that we can apply $O_P^1$ on $T$  to produce an output table $O_P^1(T)$. We then iterate, and at step (2) we feed $O_P^1(T)$ as the new input table into our model, to predict the most likely $O_P^2(T)$, and produce an output table $O_P^2(O_P^1(T))$. This iterates until at the $k$-th step, a ``\code{none}'' transformation is predicted  (recall that ``\code{none}'' is a no-op operator in our DSL in Table~\ref{tab:dsl}, to indicate that the input table is already relational and requires no transformations). The resulting $M = (O_P^1, O_P^2, \ldots)$ then becomes the multi-step transformations we synthesize for the original $T$.

The procedure above is an intuitive sketch of multi-step synthesis, though it considers only the top-1 choice at each step. In general we need to consider top-k choices at each step, to find the most likely multi-step transformations overall. We perform the general search procedure of the most likely top-$k$ steps using beam search~\cite{murphy2012machine}, as outlined in Algorithm~\ref{alg:multistep_pipeline}.

%we can intuitively choose the top-$k$  candidate operators with the highest probability scores for the current input table. Using each of the candidate operator, we can perform a single-step transformation to obtain a candidate transformed table. However, real-world tables sometimes contain multiple issues and the table after a single-step transformation may still be non-relational. Therefore, we need to run our model again on each candidate transformed table to generate $k$ candidate operators for the second-step transformation. We can repeat this process until the table is relationalized. Since our model will predict a ``none" operator when the table is relationalized, we can use it as the stopping criteria to automate the process. That is to say, if the candidate operator is  a ``none" operator, we stop further transformations on the table. 

We start in Algorithm~\ref{alg:multistep_pipeline} with an empty pipeline $M$ and the original input table $T$.  At each iteration, we invoke model $H$ on top-$k$ output tables from the last iteration, to obtain the top $k$ candidate operators for each (Line 6). We perform the predicted transformations and expand each $M$ with one additional predicted step to get $M_{next}$ (Line 8), whose probability can be computed as the product of the probability of its operators (Line 9).
 If a predicted operator is ``\code{none}", we reach a terminal state and save it as a candidate pipeline (Line 10-11). Otherwise, we keep the current pipeline in the beam for further search (Line 13).
At the end of each iteration, we rank all partial pipelines by probabilities, and keep only the top $k$ pipelines with the highest probability (Line 14).  We terminate the search after a total of $L$ steps (Line 3), and return the top-$k$  with the highest probabilities as output (Line 15-16).
%In our experiments, we observe that most tables can be relationalized with only one or two steps. As a result, we set the length limit to be $L=3$ in our experiments.  
%Finally, we return the top k candidate pipelines with the highest probability scores (Line 17-18). 

We demonstrate  Algorithm~\ref{alg:multistep_pipeline} using the following example.

\begin{small}
\begin{algorithm}[t]
\SetKw{kwReturn}{return}
 \Input{\at model $H$, input table $T$} %, number of synthesized pipelines $k$}
 \Output{Top-$k$ predicted pipelines by probabilities: $M_1$, $M_2$ ... $M_k$}
 $Cands = []$, 
 $M \leftarrow []$, 
 $M.prob = 1$ \tcp{initialize} 
 $B_{cur} \leftarrow [(T, M)]$ \\

 \For{i = 1, 2, ... L}{
    $B_{next} \leftarrow []$ \\
    \ForEach{$(T, M)$ in $B_{cur}$}{
       $\hat{O_{p1}}, \hat{O_{p2}} .,. \hat{O_{pk}}  \leftarrow H(T)$ 
       \tcp{top k predictions}
       \For{j = 1, 2, ...k}{
            $T_{next} \leftarrow \hat{O_{pj}} (T)$, $M_{next} \leftarrow M.append(\hat{O_{pj}})$\\
            $M_{next}.prob \leftarrow M.prob \times \hat{O_{pj}}.prob$ \\
            \If{$\hat{O_{pj}} = none$}{
                $Cands.append(M_{next})$
            }
            \Else{
                $B_{next}.append((T_{next}, M_{next}))$
            }
       }
    }
    sort $B_{next}$ by $M.prob$, $B_{cur} \leftarrow B_{next}[:k]$\\
    
 }

 Sort $Cands$ by $M.prob$ \\
\kwReturn $Cands[:k]$
\caption{Multi-step pipeline synthesis by top-k search}
\label{alg:multistep_pipeline}
\end{algorithm}
\end{small}

\begin{example}
\label{ex:algo}
We revisit Example~\ref{ex:multi-step}.  Given the input table $T$ shown on the left of Figure~\ref{fig:multi-step-ex}, we invoke our trained model $H$ to predict likely transformations, where the top-2 is: (1) $O_1$: ``\code{transpose}'' with probability 0.5, which leads to an output table $O_1(T)$ (shown in the middle of Figure~\ref{fig:multi-step-ex}), (2) $O_2$: ``\code{stack}'' (with parameters: start-idx = Col-B, end-idx=Col-E) which also has a probability 0.5, that will lead to an output table $O_2(T)$. We keep both 1-step candidates $\{O_1, O_2\}$, and continue our search of possible second steps. 

For the second step, if we follow the path of $O_1$ we will operate on $O_1(T)$ as the new input table, for which the top-2 predicted steps is: (1) $O_3$ ``\code{stack}'' (start-idx = Col-C, end-idx=Col-E), with probability 0.8, and (2) $O_4$ ``\code{none}'' with probability 0.1. Alternatively, if we follow the path of $O_2$ we would have $O_2(T)$ as the new input, for which we also generate its top-2. This leads to a total of $2 \times 2=4$ possible 2-step transformations, from which we pick the top-2 with the highest probabilities, to continue our search with 3-steps, etc.

We rank all resulting multi-step transformations by probabilities. This returns $\{O_1, O_3\}$ as the most likely (with probability 0.5*0.8 = 0.4), which is indeed the desired transformation in Example~\ref{ex:multi-step}.
\end{example}
\vspace{-2mm}

\vspace{-1mm}
\subsection{Input/output Re-ranking}
\label{subsec:rerank}

So far, our synthesis model is ``input-only'', as it only uses the characteristics of the input table $T$ to predict transformations $M$. However, sometimes this is not enough, as the characteristics of the output table, $M(T)$ would also provide useful signals. We illustrate this using the following example.

\begin{example}
\label{ex:rerank}
In Example~\ref{ex:algo}, based only on the input $T$ in Figure~\ref{fig:multi-step-ex}, our model predicts both $O_1$  ``\code{transpose}'' and $O_2$  ``\code{stack}'' as possible choices (both with probability=0.5).  ``\code{Stack}'' was incorrectly ranked high, because from $T$ alone ``\code{stack}'' looks plausible, as $T$ has a large number of homogeneous columns (Col-B to E), which fits the typical pattern for ``\code{stack}'' as shown in Figure~\ref{fig:combined-ex}(a).

We can better predict whether $O_1$ or $O_2$ is more suitable, if we apply both programs on $T$ and inspect the resulting output $O_1(T)$ and $O_2(T)$. It can be verified that for $O_1(T)$ values in the same columns are homogeneous, whereas $O_2(T)$ (using ``\code{stack}'') leads to a table where values such as ``\code{ES}'', ``\code{MS}'' (from ``\code{GroupID}'')  become intermixed with integers in the same columns, which is not homogeneous and not ideal, and is something that our tabular model can detect and penalize. Inspecting the output $O_1(T)$ and $O_2(T)$ thus allows us to correctly re-rank $O_1$ as a more likely transformation than $O_2$, which is difficult when a model looks at $T$ alone.

%into our re-ranking model, whose embedding/feature-extraction layers recognize that the output $O_2(T)$ does not look ideal, because after ''\code{stack}'', $O_2(T)$ would have ``\code{GroupID}'' values (``\code{ES}'', ``\code{MS}'', etc.) intermix with integer numbers in the same column, which is not a desirable relational form and thus penalized by re-ranking that has access to both input and output tables. After re-ranking, we get $P(O_1) = 0.8$  and $P(O_2) = 0.2$, making it easier to identify the correct transformation.

%Given the input $T$ in Figure~\ref{fig:multi-step-ex}, in Example~\ref{ex:algo} our model predicts both $O_1$: ``\code{transpose}'' and $O_2$: ``\code{stack}'' to be likely with probability=0.5, because $T$ has a large number of homogeneous columns, making ``\code{stack}'' a plausible choice based on $T$ alone. 

%Using input/output re-ranking, we feed the output from both candidates $O_1(T)$ and $O_2(T)$ into our re-ranking model, whose embedding/feature-extraction layers recognize that the output $O_2(T)$ does not look ideal, because after ''\code{stack}'', $O_2(T)$ would have ``\code{GroupID}'' values (``\code{ES}'', ``\code{MS}'', etc.) intermix with integer numbers in the same column, which is not a desirable relational form and thus penalized by re-ranking that has access to both input and output tables. After re-ranking, we get $P(O_1) = 0.8$  and $P(O_2) = 0.2$, making it easier to identify the correct transformation.
\end{example}
% \vspace{-2mm}

%However, we observe that in some cases, it may be hard to determine whether an operator is suitable based on  the input table, but it is much easier to determine whether the transformed table induced by the operator is relational or not. \peng{examples: xxx}. Also, as we will show in the experiments, sometimes our model can include the ground truth operator in the top $k$ predictions, but its ranking is not high. Therefore, to improve the performance of our model, our idea is to rerank the predicted operators based on the  output transformed tables. 

\begin{figure}[t]
%\vspace{-3mm}
    \centering
    \includegraphics[width=\columnwidth]{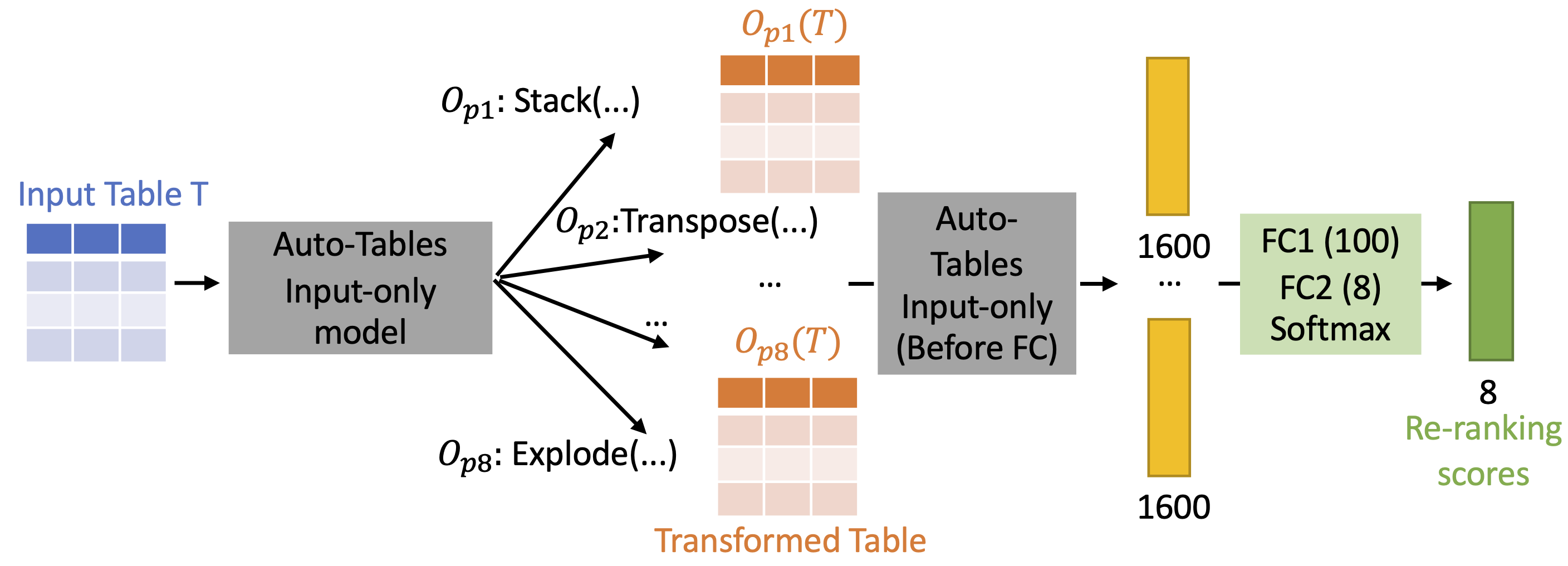}
    \vspace{-8mm}
    \caption{Input/output re-ranking: model architecture.}
    \label{fig:reranking_model_arch}
\vspace{-5mm}
\end{figure}

This motivates us to develop an ``input/output-based'' re-ranking model as shown in Figure~\ref{fig:reranking_model_arch}. After the input-only synthesis model (Section~\ref{subsec:input-model}) produces top-$k$ likely operators $\{O_{pi}, i \in [k]\}$ (e.g., we consider top-8 operators for re-ranking in our experiments), the re-ranking model will look at all output transformed tables $\{O_{pi}(T), i \in [k]\}$ and aims to generate a re-ranking score for each of them indicating which operator is more suitable based on the output transformed tables. To do so, similar to the input-only model, we need to first convert each transformed table into a feature vector using table embedding, dimension reduction and feature extraction layers. Since the input-only model has been trained well at this time, we directly reuse the architecture and weights of these layer from the pre-trained input-model~\footnote{We remove the fully-connected output layers from the input model, which are specific to predicting synthesis outcomes and not relating to extracting table features.}. We then concatenate the feature vectors of all transformed tables and use fully connected layers followed by a softmax function to produce a $k$-dimension vector as re-ranking scores. For training, we consider the re-ranking as a classification task to predict which of the $k$ transformed tables is the ground truth. Thus, the training loss can be computed using cross-entropy loss. We train the re-ranking model using the same training data generated from self-supervision in Section~\ref{sec:training_data_generation}

%% file: experiment.tex
\section{Experiments}
\label{sec:exp}
We perform extensive evaluation on the performance of different algorithms, using real test data. The results show that our method significantly outperforms the baseline methods in terms of both quality and efficiency. Our labeled benchmark data is available on GitHub\footnote{\url{https://github.com/LiPengCS/Auto-Tables-Benchmark}} for future research.

% \kr{can you summarize the high level takeaways here}

\subsection{Experimental Setup}

\stitle{Benchmarks.} To study the performance of our method in real-world scenarios, we compile an \atbench benchmark using real cases from three sources: (1) online user forums, (2) Jupyter notebooks, and (3) real spreadsheet-tables and web-tables.

\underline{Forums.} Both technical and non-technical users ask questions on forums, regarding how to restructure their tables. As  Figure~\ref{fig:stackoverflow-ex} shows, users often provide sample input/output tables to demonstrate their needs.  We sample 23 such questions from StackOverflow and Excel user forums as test cases. (We feed \at with user-provided input tables, and evaluate whether the correct transformation can be synthesized to produce the desired output table given by users).

\underline{Notebooks.} Data scientists frequently restructure tables using Python Pandas, often inside Jupyter Notebooks. We sample 79 table-restructuring steps extracted from the Jupyter Notebooks crawled in~\cite{yan2020auto, yang2021auto} as our test cases. We use the transformations programmed by data scientists as the ground truth.

\underline{Excel+Web.} A large fraction of tables ``in the wild'' require transformations before they are fit for querying, as shown in Figure~\ref{fig:combined-ex} and~\ref{fig:combined-web-ex}. We sample 56 real web-tables and 86 spreadsheet-tables (crawled from a search engine) that require such transformations, and manually write the desired transformations as the ground truth.

Combining these sources, we get a total of 244 test cases as our \atbench (of which 26 cases require multi-step transformations). Each test case consists of an input table $T$, ground-truth transformations $M_g$\footnote{It should be noted that for some test cases, there may be more than one transformation sequence that can produce the desired output. We  enumerate all such sequences in our ground-truth, and mark an algorithm as correct as long as it can synthesize one ground-truth sequence.}, and an output table $M_g(T)$ that is relational. 
\iftoggle{fullversion}
{
}
{
Detailed statistics of the benchmark can be found in our technical report~\cite{full}.
}

\iftoggle{fullversion}
    {
    \begin{table}[t]
    %\vspace{-8mm}
    \caption{Details of \atbench Benchmark}
    \vspace{-2mm}
    \label{tab:autotable_benchmark}
    \scalebox{0.8}{
    \begin{tabular}{l|cccc|c}
    \toprule
    \textbf{} & \textbf{Forum} & \textbf{Notebook} & \textbf{Excel} & \textbf{Web} & \textbf{Total} \\
    \midrule
    \textbf{Single-Step} & \textbf{23} & \textbf{75} & \textbf{65} & \textbf{55} & \textbf{218} \\
    \quad{} - transpose & 0 & 11 & 11 & 6 & 28 \\
    \quad{} - stack & 10 & 20 & 2 & 24 & 56 \\
    \quad{} - wtl & 6 & 24 & 1 & 3 & 34 \\
    \quad{} - explode & 2 & 17 & 14 & 15 & 48 \\
    \quad{} - ffill & 0 & 0 & 11 & 7 & 18 \\
    \quad{} - pivot & 5 & 3 & 0 & 0 & 8 \\
    \quad{} - subtitle & 0 & 0 & 26 & 0 & 26 \\
    \textbf{Multi-Step} & \textbf{0} & \textbf{4} & \textbf{21} & \textbf{1} & \textbf{26} \\
    \midrule
    \textbf{Total} & \textbf{23} & \textbf{79} & \textbf{86} & \textbf{56} & \textbf{244} \\
    \bottomrule
    \end{tabular}
    }
    \end{table}
}
{
}

% results table moved up

\begin{table*}[!h]
    \vspace{-15mm}
  \begin{minipage}{\columnwidth}
    \centering
    \caption{Quality comparison using Hit@k, on 244 test cases}
    \vspace{-4mm}
    \label{tab:hit_at_k_pos} 
    \resizebox{\columnwidth}{!}{%
        \begin{tabular}{ccccccccc}
        \toprule
        \multirow{2}{*}{\textbf{Method}} & \multicolumn{4}{c}{\textbf{No-example methods}} & \multicolumn{4}{c}{\textbf{By-example methods}} \\ \cmidrule(lr){2-5} \cmidrule(lr){6-9}
         & \textbf{Auto-Tables} & \textbf{TaBERT} & \textbf{TURL} & \textbf{GPT-3.5-fs} & \textbf{FF} & \textbf{FR} & \textbf{SQ} & \textbf{SC} \\
         \midrule
        Hit @ 1 & \textbf{0.570} & 0.193 & 0.029  & 0.196  &  0.283 &  0.336 & 0 & 0 \\
        Hit @ 2 & \textbf{0.697} & 0.455 & 0.071 & - & - & - & 0 & 0 \\
        Hit @ 3 & \textbf{0.75} & 0.545  &  0.109 & - &  - & - & 0 & 0 \\
        Upper-bound & - & - & - & - &  0.471 & 0.545  &  0.369 &  0.369 \\
        \bottomrule
        \end{tabular}
    }
    \end{minipage}\hfill % maximize the horizontal separation
    \begin{minipage}{\columnwidth}
    \centering
    \caption{Synthesis latency per test case }
    \vspace{-4mm}
    \label{tab:running_time}
    \resizebox{\columnwidth}{!}{%
        \begin{tabular}{cccc}
        \toprule 
        \textbf{Method} & \textbf{Auto-Tables} & 
        \textbf{\begin{tabular}[c]{@{}c@{}}Foofah \\ (excl. 110 timeout cases)\end{tabular}} & \textbf{\begin{tabular}[c]{@{}c@{}}FlashRelate\\ (excl. 91 timeout cases)\end{tabular}}  \\
        \midrule
        50 \%tile &\textbf{0.127s} & 0.287s + human effort & 3.4s + human effort  \\
        90 \%tile &\textbf{0.511s} & 22.891s + human effort & 57.16s + human effort  \\
        95 \%tile &\textbf{0.685s} & 39.188s + human effort & 348.6s + human effort  \\
        Average &\textbf{0.224s} & 5.996s + human effort & 59.194s + human effort  \\
        \bottomrule
        \end{tabular}
    }
  \end{minipage}
\end{table*}

\begin{table*}[!h]
  \begin{minipage}[t]{\columnwidth}
    \centering
    \caption{Ablation Studies of \at}
    \label{tab:ablation}
    \vspace{-4mm}
    \resizebox{\columnwidth}{!}{%
        \begin{tabular}{cccccccc}
        \toprule
        \multirow{2}{*}{\textbf{Method}} & \multirow{2}{*}{\textbf{Full}} & \multirow{2}{*}{\textbf{No Re-rank}} & \multicolumn{5}{c}{\textbf{No Re-rank \&}} \\ \cmidrule(lr){4-8} 
        
         &  &  & \textbf{No Aug} & \textbf{No Bert} & \textbf{No Syn} & \textbf{1x1 Only} & \textbf{5x5} \\
         \midrule
        Hit@1 & \textbf{0.570} & 0.508 & 0.463 & 0.467 & 0.504 & 0.471 & 0.480 \\
        Hit@2 & \textbf{0.697} & 0.652 & 0.582 & 0.627& 0.648 & 0.607 & 0.594 \\
        Hit@3 & \textbf{0.75} & 0.730 & 0.656 & 0.693 & 0.676 & 0.652 & 0.660 \\
        \bottomrule
        \end{tabular}
    }
    \end{minipage}\hfill % maximize the horizontal separation
    \vspace{-4mm}
      \begin{minipage}[t]{\columnwidth}
        \centering
        \caption{Sensitivity to different semantic embeddings. }
        \label{tab:vary_embed}
        \vspace{-4mm}
        \resizebox{\columnwidth}{!}{%
            \begin{tabular}{ccccc}
            \toprule
            \textbf{Embedding methods} & \textbf{sentenceBERT} & \textbf{fastText} & \textbf{GloVe} & \textbf{No Semantic} \\
            \midrule
            Hit@1 & 0.508 & 0.529 & 0.525 &  0.467 \\
            Hit@2 & 0.652 & 0.656 & 0.676 &  0.627\\
            Hit@3 & 0.730 & 0.734 & 0.734 &  0.734 \\  \hline
            \begin{tabular}[c]{@{}c@{}}Avg. latency per-case \\ w/ this embedding\end{tabular}
            & 0.299s & 0.052s & 0.050s & 0.026s \\
            \bottomrule
            \end{tabular}
        }
  \end{minipage}
\end{table*}

\stitle{Evaluation Metrics.} We evaluate the quality and efficiency of different algorithms in synthesizing transformations.

\underline{Quality}. Given an input table $T$, an algorithm $A$ may generate top-$k$ transformations $(\hat{M}_1, \hat{M}_2, ... \hat{M}_k)$, ranked by probabilities, for users to inspect and pick. We evaluate the success rate of synthesis using the standard $Hit@k$ metric~\cite{ir-book}, defined as:
\begin{align*}
    Hit@k(T) = \sum_{i=1}^k \mathbf{1}(\hat{M}_{i}(T) = M_g(T))
\end{align*}
which looks for exact matches between the top-$k$ ranked predictions ($\hat{M}_{i}(T), 1 \leq i \leq k$) and the ground-truth $M_g(T)$.  The overall $Hit@k$ on the entire benchmark, is then simply the average across all test cases $T$. We report $Hit@k$ up to $k=3$.

\underline{Efficiency}. We report the latency of synthesis using wall-clock time. All experiments are conducted on a Linux VM with 24 vCPU cores, and 4 Tesla P100 GPUs.

\stitle{Methods Compared.} We compare with the following methods.

\begin{itemize}[leftmargin=*]
\item \textit{\at}. This is our approach and is the only method that does not require users to provide input/output examples (unlike other existing methods). In order to train \at, we generate 1.4M (input-table, transformation) pairs evenly distributed across 8 operators, following the self-supervision procedure  (Section~\ref{sec:training_data_generation}), using 15K base relational tables crawled from public sources~\footnote{We use the dataset from~\cite{lin2023auto}, which has thousands of relational Power-BI models crawled from public sources. We sample 15K fact and dimension tables from these models as our ``base'' relational tables. %Details of our data pre-processing can be found in our technical report~\cite{full} in the interest of space. 
Since our training data is collected via Power-BI data models, they are completely separate from our test data (Web and Excel tables).}. We take a fixed size of input with the first 100 rows and 50 columns at the top-left corner of each table and use zero-padding for tables with less rows or columns. We implement our method using 
PyTorch~\cite{paszke2017automatic}, trained using Adam optimizer, with a learning rate of 0.001 for 50 epochs, using a batch size of 256. 

%To ensure that all tables in a batch have the same size, we crop/pad the input table to have $100$ rows and $50$ columns.

% \kr{can you briefly comment on how similar/different is the training tables from the test case tables? e.g., average number of rows, columns etc.}

\item \textit{Foofah (FF)}~\cite{jin2017foofah} synthesizes transformations based on input/output examples. We use 100 cells from the top-right of the ground-truth output table for Foofah to synthesize programs, which simulate the scenario where a user types in 100 output cells (a generous setting as it is unlikely that users are willing to provide so many examples in practice).
We test Foofah using the authors original implementation~\cite{foofah-code}, and we time-out each case after 30 minutes. 

\item \textit{Flash-Relate (FR)}~\cite{barowy2015flashrelate} is another approach to synthesize table-level transformations, which however would require input/output examples. We used an open-source re-implementation of FlashRelate~\cite{flash-relate-code} (since the original system is not publicly available), and we provide it with 100 example output cells from the ground-truth. We use  a similar time-out of 30 minutes for each test case.

\item \textit{SQLSynthesizer (SQ)}~\cite{SQLSynthesizer} is a SQL-by-example algorithm that synthesizes SQL queries based on input/output examples. We use the authors implementation~\cite{Scythe-code}, provide it with 100 example output cells, and also set a time-out of 30 minutes.

\item \textit{Scythe (SC)}~\cite{sql-by-example} is another SQL-by-example method. We used the author's implementation~\cite{PATSQL-code} and provide it with 100 example output cells, like previous methods.
%\item \textit{QBO}~\cite{qbo} \yeye{All 3 Sql-by-example work should technically have 0 coverage of the tasks we have. Though can still try experimentally with Scythe to see.}

\item \textit{TaBERT}~\cite{yin2020tabert} is a table representation approach developed in the NLP literature, and pre-trained using table-content and captions for NL-to-SQL tasks. To test the effectiveness of TaBERT in our transformation task, we replace the table representation in \at (i.e., output of the feature extraction layer in Figure~\ref{fig:input_model_arch}) with TaBERT's representation, and  train the following fully connected layers using the same training data as ours.

% We replace our table representation in \at (Section~\ref{subsec:input-model}) with TaBERT's representation, to test the effectiveness of TaBERT in our transformation task.

\item \textit{TURL}~\cite{deng2022turl} is another table representation approach for data integration tasks. Similar to \textit{TaBERT}, we test the effectiveness of TURL by replacing \at representation with  TURL's.

\item \textit{GPT}~\cite{gpt} is a family of large language models pre-trained on text and code, which can follow instructions to perform a variety of tasks. While we do not expect GPT to perform well on \at tasks, we perform a comparison nevertheless, using GPT-3.5\footnote{We used the ``gpt-3.5-turbo'' API endpoint, accessed in July 2023.} as a baseline. We perform few-shot in-context learning, using a description of the operators, together with pairs of (input-table, desired-operator) in the prompt to demonstrate the task. We provide one example demonstration per operator, for a total of 7 examples (which fit in the context allowed by GPT-3.5). We denote this method as GPT-3.5-fs (few-shot).\footnote{Note that GPT-3.5-fs is still a no-example method, as we use general-purpose examples to demonstrate each operator in our few-shot examples, which are fixed and do not vary based on different input tables.}
\end{itemize}

%\stitle{Training Process.} We implement our method using PyTorch~\cite{paszke2017automatic}. We train our model using a Adam optimizer with learning rate 0.001 for 50 epochs. The batch size is set to be 256. To ensure that all tables in a batch have the same size, we crop/pad the input table to have $100$ rows and $50$ columns.

% Since we are more interested in the performance of our method on non-relational tables, we also compute the Hit@k score only using non-relational tables and we name it as \textit{Positive Hit@k}, defined as follows.

% $$Pos \ Hit@k = \frac{\sum_{i=1}^n\sum_{j=1}^k \mathbb{1}(y_{ij} == y_i \land y_i \neq \text{None})}{\sum_{i=1}^n \mathbb{1}(y_i \neq \text{None})}$$

% We first compare the end-to-end performance of different methods. In this case, we run our model until the predicted operator is None.

%\subsection{Overall Comparisons}

\subsection{Experiment Results}

% % \begin{table}[!h]
% \begin{table}[t]
% \caption{Quality comparison using Hit@k, on 194 test cases}
% \label{tab:hit_at_k_pos} 
% \vspace{-3mm}
% \scalebox{0.75}{
% \begin{tabular}{cccccccc}
% \toprule
% \multirow{2}{*}{\textbf{Method}} & \multicolumn{3}{c}{\textbf{No-example methods}} & \multicolumn{4}{c}{\textbf{By-example methods}} \\ \cmidrule(lr){2-4} \cmidrule(lr){5-8}
%  & \textbf{Auto-Tables} & \textbf{TaBERT} & \textbf{TURL} & \textbf{FF} & \textbf{FR} & \textbf{SQ} & \textbf{SC} \\
%  \midrule
% Hit @ 1 & \textbf{0.562} & 0.187 & 0.027 & 0.285 & 0.351 & 0 & 0 \\
% Hit @ 2 & \textbf{0.68} & 0.43 & 0.075 & - & - & 0 & 0 \\
% Hit @ 3 & \textbf{0.722} & 0.539 & 0.124 & - & - & 0 & 0 \\
% Upper-bound & - & - & - & 0.421 & 0.538 & 0.34 & 0.34 \\
% \bottomrule
% \end{tabular}
% }
% \end{table}
% % \vspace{-2em}
% \begin{table}[t]
% \caption{Synthesis latency per test case }
% \label{tab:running_time}
% \vspace{-3mm}
% \scalebox{0.75}{
% \begin{tabular}{cccc}
% \toprule 
% \textbf{Method} & \textbf{Auto-Tables} & 
% \textbf{\begin{tabular}[c]{@{}c@{}}Foofah \\ (excl. 89 timeout cases)\end{tabular}} & \textbf{\begin{tabular}[c]{@{}c@{}}FlashRelate\\ (excl. 91 timeout cases)\end{tabular}}  \\
% \midrule
% 50 percentile &\textbf{0.168s} & 0.326s + human effort & 3.4s + human effort  \\
% 90 percentile &\textbf{0.553s} & 27.146s + human effort & 57.16s + human effort  \\
% 95 percentile &\textbf{0.690s} & 41.411s + human effort & 348.6s + human effort  \\
% Average &\textbf{0.299s} & 6.995s + human effort & 59.194s + human effort  \\
% \bottomrule
% \end{tabular}
% % \vspace{-3mm}
%  }
% \end{table}

\begin{figure*}
\vspace{-19mm}
\begin{minipage}[b]{0.33\textwidth}
   \centering
{\includegraphics[width=0.9\textwidth]{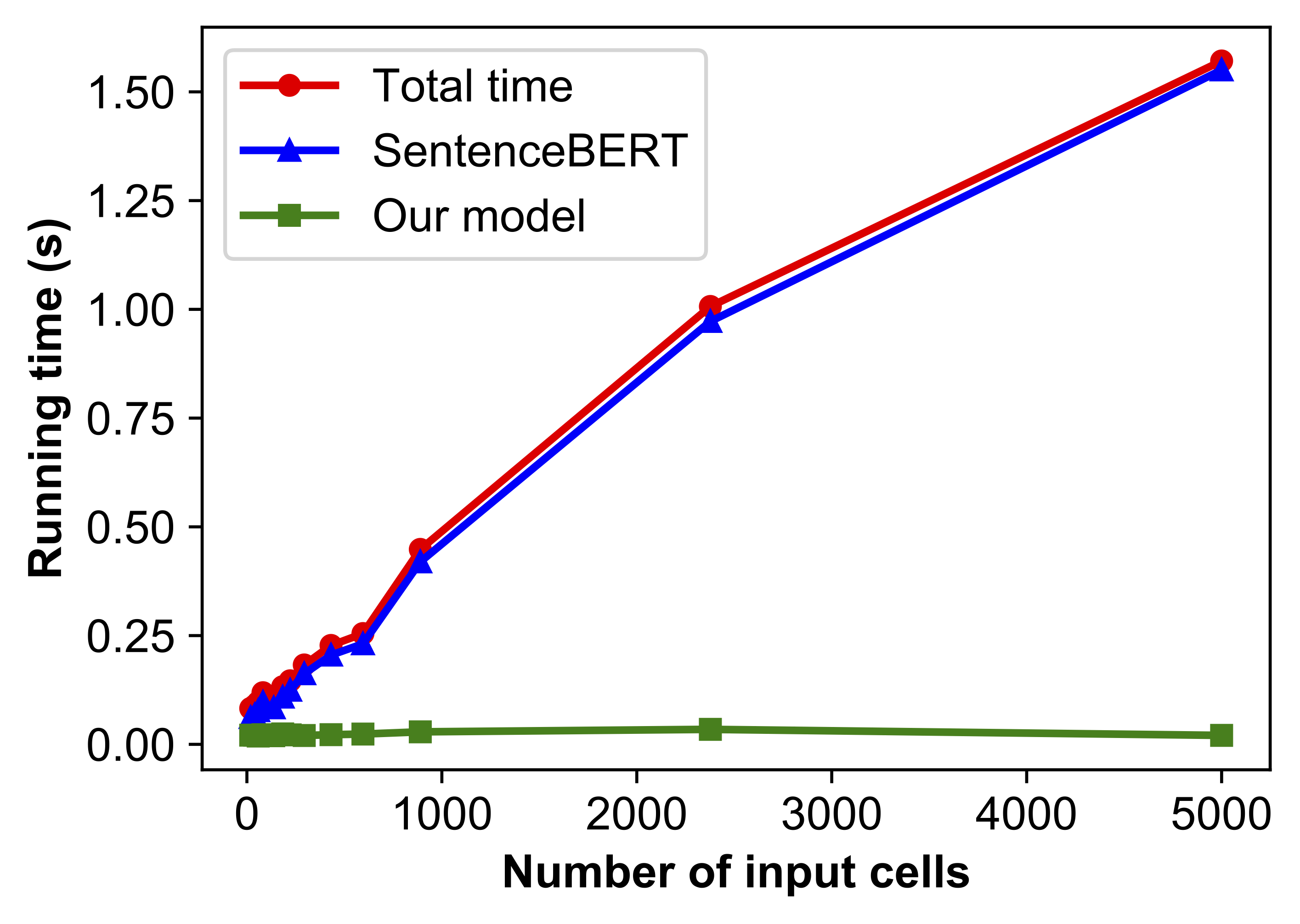}}
\vspace{-5mm}
\caption{\at latency analysis}
\label{fig:running_time}
\end{minipage}\hfill
\begin{minipage}[b]{0.33\textwidth}
   \centering    {\includegraphics[width=0.9\textwidth]{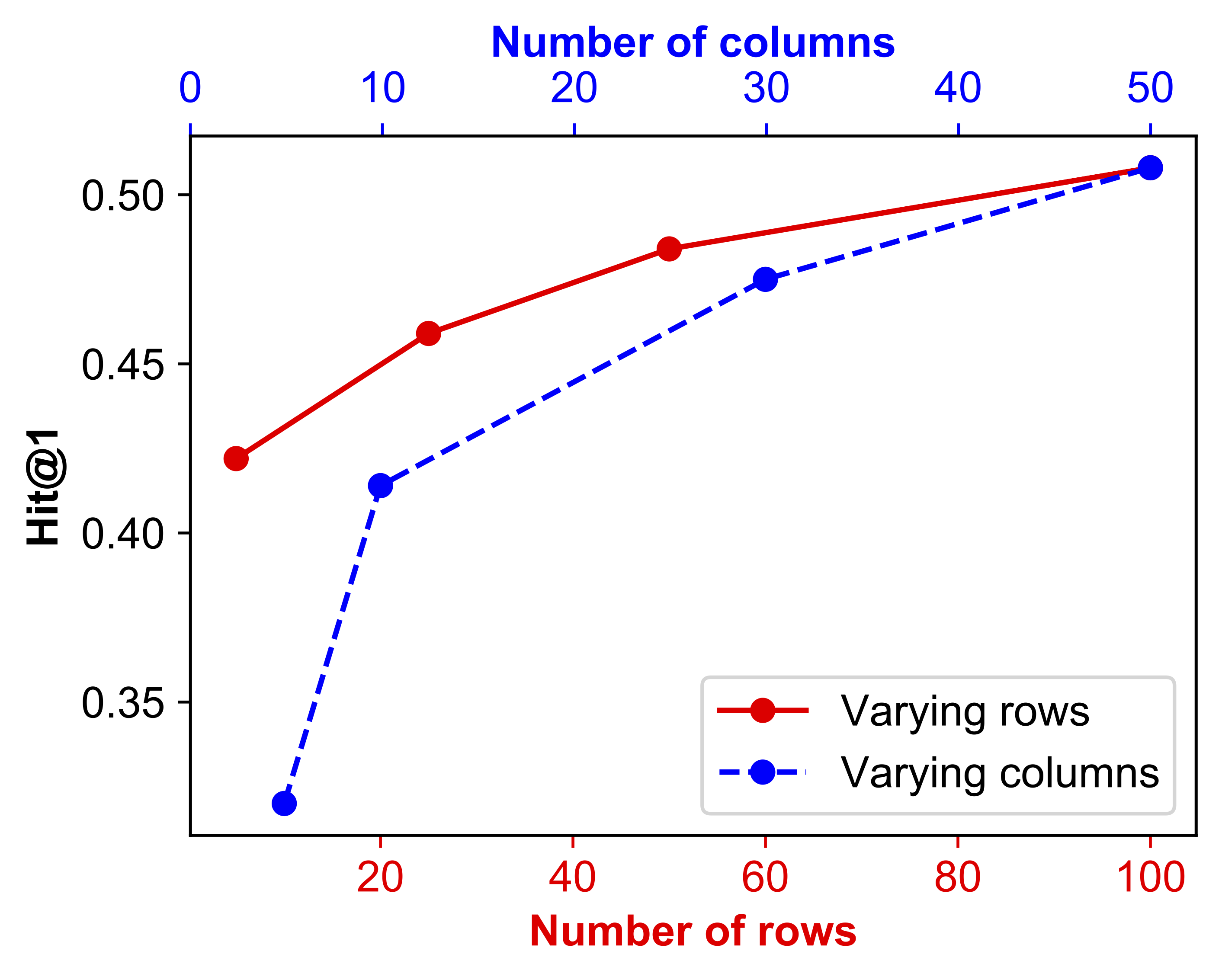}}
\vspace{-5mm}
\caption{Vary input size}
\label{fig:vary_input}
\end{minipage} \hfill
\begin{minipage}[b]{0.33\textwidth}
   \centering
{\includegraphics[width=0.9\textwidth]{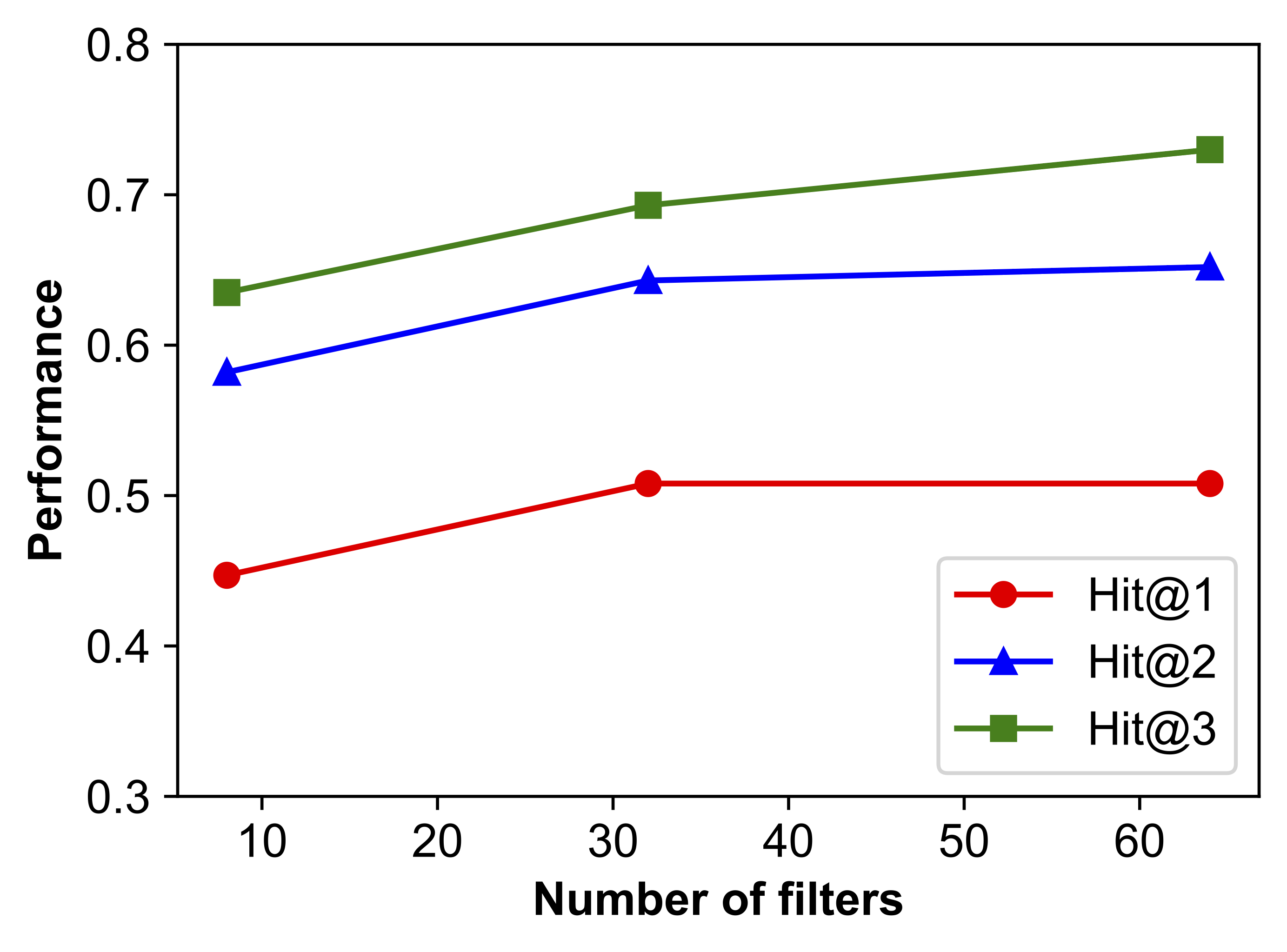}}
\vspace{-5mm}
\caption{Vary number of filters}
\label{fig:vary_filter}
\end{minipage} \hfill
\vspace{-3mm}
\end{figure*}

\stitle{Quality Comparison.}  Table~\ref{tab:hit_at_k_pos} shows the comparison between \at and baselines, evaluated on our benchmark with 244 test cases. We group all methods into two classes:  (1) ``No-example methods'' that do not require users to provide any input/output examples, which include our \at, and variants of \at that use TaBERT and TURL for table representations, respectively; and (2) ``By-example methods'' that include Foofah (FF), FlashRelate (FR), SQLSynthesizer (SQ), and Scythe (SC), all of which are provided with 100 ground truth example cells. 

As we can see, \at significantly outperforms all other methods, successfully transforming 75\% of test cases in its top-3, \textit{without needing users to provide any examples},  despite the challenging nature of our tasks. Recall that in our task, even for a single-step transformation, there are thousands of possible operators+parameters to choose from (e.g., a table with 50 columns that requires ``\code{stack}'' will have 50x50 = 2,500 possible parameters of start\_idx and end\_idx) and for two-step transformations, the search space is in the millions (e.g., for ``\code{stack}'' alone it is $2500^2 \approx 6M$), which is clearly non-trivial. 

% This is despite our task being very challenging -- the search space for 2-step transformations alone amounts to $270^2 \approx 72K$ possible choices (recall that our output vector has 270 dimensions), which makes this clearly non-trivial. 

Compared to other no-example methods, \at outperforms TaBERT and TURL respectively by 37.7 and 54.1 percentage point on Hit@1, 20.5 and 64.1 percentage point on Hit@3. This shows the strong benefits for using our proposed table representation and model architecture, which are specifically designed for the table transformation task (Section~\ref{subsec:input-model}). 

% Compared to TaBERT and TURL, we see strong benefits (30-50 percentage points) for using our proposed table representation and model architecture, which are specifically designed for the table transformation task (Section~\ref{subsec:input-model}). 

Compared to by-example methods, the improvement of \at is similarly strong. Considering the fact that these baselines use 100 output example cells (which users need to manually type), whereas our method uses 0 examples, we argue that \at is clearly a better fit for the table-restructuring task at hand. Since some of these methods (FF and FR) only return top-1 programs, we also report in the last row their ``upper-bound'' coverage, based on their DSL (assuming all transformations supported in their DSL can be successfully synthesized).

\iftoggle{fullversion}
{

    \underline{Result breakdown by benchmark sources.} We additionally drill down on our quality results, using a break-down by benchmark data sources (forum, notebooks, and Excel+web). We report in Table~\ref{tab:quality-breakdown} the performance of three best-performing methods: \at (AT), TabERT (TA), FlashRelate (FR), in the interest of space.  

    It can be seen from the table that the quality of \at is consistent across the three, confirming its effectiveness across diverse test cases arising from different sources.

    %     \begin{table}[t]
    %     \caption{Quality comparisons, broken down by data sources}
    %     \label{tab:quality-breakdown}
    %     \scalebox{0.8}{
    %     \begin{tabular}{cccccccccc}
    %     \toprule
    %      & \multicolumn{3}{c}{\textbf{Forum}} & \multicolumn{3}{c}{\textbf{Notebook}} & \multicolumn{3}{c}{\textbf{Excel+Web}} \\ \cmidrule(lr){2-4}\cmidrule(lr){5-7}\cmidrule(lr){8-10}
    %     \textbf{Method} & \textbf{AT} & \textbf{TA} & \textbf{FR} & \textbf{AT} & \textbf{TA} & \textbf{FR} & \textbf{AT} & \textbf{TA} & \textbf{FR} \\ \midrule
    
    % Hit @ 1 & \textbf{0.521} & 0.217 & 0.043 & \textbf{0.582} & 0.215 & 0.278 & \textbf{0.554} & 0.154 & 0.489 \\
    % Hit @ 2 & \textbf{0.696} & 0.522 & - & \textbf{0.722} & 0.531 & - & \textbf{0.641} & 0.319 & - \\
    % Hit @ 3 & \textbf{0.696} & 0.565 & - & \textbf{0.747} & 0.633 & - & \textbf{0.706} & 0.451 & -\\
        
    %     \bottomrule
    %     \end{tabular}
    %     }
    %     \end{table}
    \begin{table}[t]
    \vspace{3mm}
    \caption{Quality comparisons by data sources}
    \vspace{-3mm}
    \label{tab:quality-breakdown}
    \resizebox{\columnwidth}{!}{
    \begin{tabular}{ccccccccccccc}
    \toprule
     & \multicolumn{3}{c}{\textbf{Forum}} & \multicolumn{3}{c}{\textbf{Notebook}} & \multicolumn{3}{c}{\textbf{Excel}} & \multicolumn{3}{c}{\textbf{Web}} \\ \cmidrule(lr){2-4}\cmidrule(lr){5-7}\cmidrule(lr){8-10}\cmidrule(lr){11-13}
    \textbf{Method} & \textbf{AT} & \textbf{TA} & \textbf{FR} & \textbf{AT} & \textbf{TA} & \textbf{FR} & \textbf{AT} & \textbf{TA} & \textbf{FR} & \textbf{AT} & \textbf{TA} & \textbf{FR} \\
    \midrule
    Hit @ 1 & \textbf{0.522} & 0.217 & 0.043 & \textbf{0.582} & 0.241 & 0.278 & \textbf{0.558} & 0.174 & 0.5 & \textbf{0.589} & 0.143 & 0.286 \\
    Hit @ 2 & \textbf{0.696} & 0.478 & - & \textbf{0.722} & 0.557 & - & \textbf{0.651} & 0.442 & - & \textbf{0.732} & 0.321 & - \\
    Hit @ 3 & \textbf{0.696} & 0.565 & - & \textbf{0.747} & 0.62 & - & \textbf{0.709} & 0.547 & - & \textbf{0.839} & 0.429 & - \\
    \bottomrule
    \end{tabular}
    }
    \end{table}
}
{
    \underline{Additional quality results.} We report additional results on quality, such as  a breakdown by benchmark sources, and Hit@K in the presence of input tables that are already relational (for which \at should correctly detect and not over-trigger, by performing no transformations), in our technical report~\cite{full}.
}

% removed during revision, add back to full version
%\revised{}
%\begin{comment}
\iftoggle{fullversion}
{
    \underline{Quality comparisons in the presence of relational tables.}  Recall that since \at can detect input tables that are already relational, and predict ``\code{none}'' for such tables, an additional use case of \at is to invoke it on all input tables encountered in spreadsheets or on the web, which include both relational tables (requiring no transformations) and non-relational tables (requiring transformations), such that any tables that \at predicts to require transformations can then be surfaced to users to review and approve. Note that this is a use case that by-example baselines cannot support, as they require users to first manually scan and identify tables requiring transformations.

    For this purpose, we test \at on the 244 test cases that require transformations, as well as the corresponding 244 output tables that are already relational and require no transformations. Using this collection of 488 cases, we not only test whether \at can correctly synthesize transformations on non-relational input, but also whether it can correctly predict ``\code{none}'' on the relational tables not requiring transformations, using the same $Hit@K$. 
 
    Table~\ref{tab:hit_at_k} shows \at achieves high quality, suggesting that it does not ``over-trigger'' on tables that are already relational, and can be effective at this task.
    
    % \begin{table}[t]
    % \vspace{-2mm}
    % \scalebox{0.8}{
    % \begin{tabular}{cccc}
    % \toprule
    % \textbf{Method} & \textbf{Auto-Tables} & \textbf{TaBERT} & \textbf{TURL} \\
    % \midrule
    % Hit @ 1 & \textbf{0.667} & 0.276 & 0.176 \\
    % Hit @ 2 & \textbf{0.781} & 0.584 & 0.392 \\
    % Hit @ 3 & \textbf{0.829} & 0.702 & 0.454 \\
    % \bottomrule
    % \end{tabular}
    % }
    % \caption{Quality comparison using Hit@k, on 194 cases with non-relational input (requiring transformations), and 194 cases with relational input (not requiring transformations).}
    % \label{tab:hit_at_k}
    % \end{table}

    \begin{table}[t]
    \vspace{5mm}
    \caption{Quality comparison using Hit@k, on 244 cases with non-relational input (requiring transformations), and 244 cases with relational input (not requiring transformations).}
    \vspace{-2mm}
    \scalebox{0.8}{
    \begin{tabular}{cccc}
    \toprule
    \textbf{Method} & \textbf{Auto-Tables} & \textbf{TaBERT} & \textbf{TURL} \\
    \midrule
    Hit @ 1 & \textbf{0.695} & 0.258 & 0.175 \\
    Hit @ 2 & \textbf{0.803} & 0.594 & 0.387 \\
    Hit @ 3 & \textbf{0.840} & 0.699 & 0.444 \\
    \bottomrule
    \end{tabular}
    }
    \label{tab:hit_at_k}
    \end{table}
}
{

}

\stitle{Running Time.} Table~\ref{tab:running_time} compares the average and 50/90/95-th percentile latency, of all methods to synthesize one test case. \at is interactive with sub-second latency on almost all cases, whose average is 0.224.  Foofah and FlashRelate take considerably longer to synthesize, even after we exclude cases that time-out after 30 minutes. This is also not counting the time that users would have to spend typing in output examples for these by-example methods, which we believe make \at substantially more user-friendly for our transformation task.  

%to make predictions on each table and takes less than 0.69s on 95\% tables,  which can be used for real-time prediction. In comparison, Foofah and Flashrelate exceed the time limit (i.e., 30 minutes) on 89 and 91 out of 194 test cases, respectively. Excluding the timeout cases, Foofah and Flashrelate respectively take on average 6.995s and 59.194s on each table. In addition, using by-example methods (e.g., Foofah, Flashrelate), users need to provide ground truth output cells (e.g., up to 100 cells in our experiments), which can take considerable time. 

%We sort the test cases by the number of input cells and group every 20 cases as a bin. 
Figure~\ref{fig:running_time} shows the average latency of \at, on cases with different number of non-empty input cells. As we can see, the latency grows linearly as the number of cells  increases, but since we only need to use at most the top-left 100 rows and 50 columns to correctly synthesize a program, this is always bounded by a couple of seconds at most. Furthermore, we notice that the running time is dominated by SentenceBERT embedding, which accounts for 91.5\% of the latency. 
In comparison, the actual inference time of \at (the green line) is very small and almost constant.

\stitle{Ablation Study}
We perform ablation studies to understand the benefit of \at components, which is shown in Table~\ref{tab:ablation}.

\underline{Contribution of Input/Output Re-Ranking.} To study the contribution of our re-ranking model (Section~\ref{subsec:rerank}), we compare the performance of \at with and without re-ranking. Table~\ref{tab:ablation} shows that our ``Full'' method (with re-ranking) produces substantially better Hit@1 and Hit@2 compared to ``No Re-rank''. %However, there is no substantial difference for Hit@3. This is because the good options are already exhausted at top-3 positions, making the resulting difference small.

% \stitle{Contribution of Input/Output Re-Ranking.} To understand the contribution of our re-ranking (Section~\ref{subsec:rerank}), we evaluate the performance of \at with and without re-ranking. As we can see in Table~\ref{tab:ablation}, re-ranking produces substantial improvement for both Hit and Pos-Hit at the top-2 positions. There is no substantial difference for Hit@3 and Pos-Hit@3, likely because the good options are already exhausted at rank-3 positions, making the resulting difference small.

\underline{Contribution of Data Augmentation.}  To study the benefits of data augmentation in training data generation (Section~\ref{sec:training_data_generation}), we disable augmentation when generating training data (i.e., using only the base relational tables). Table~\ref{tab:ablation} shows this result under ``No Aug'', which suggests that our Hit@k drop substantially, underscoring the importance of data augmentation.

% Data augmentation increases the variation of the training set by producing multiple training examples from each relational base table .

\underline{Contribution of Embeddings.} Recall that we use both syntactic embedding and semantic embedding (sentenceBERT) to represent each cell (Section~\ref{subsec:input-model}). To understand their contributions, we remove each embedding in turn, and the results are shown under ``No Bert'' and ``No Syntactic'' in Table \ref{tab:ablation}. Both results show a substantial drop in performance, confirming their importance (semantic embedding with sentenceBERT is likely more important, as removing it leads to a more significant drop).

%two types of embeddings (No Rerank), the Hit@k generally scores become worse after dropping one of them, except that Hit@2 is slightly increased after dropping syntactic embeddings. This means both types of embeddings help \at to predict the desired transformations. Also, we can see that sentenceBERT is more important, as its removal leads to a more significant drop in overall quality. 

\underline{Contribution of 1D Filters.} Recall that we use convolution filters of size 1x1 and 1x2 to extract features from rows and columns (Section~\ref{subsec:input-model}). To understand the effectiveness of this design, we evaluate our method with alternative filters. First, we replace all the 1x2 filters with 1x1 filters. The result is labeled ``1x1 Only'' and shows a significant drop. %Especially on Hit@1, the score is dropped by 0.088 compared to the original results (No Rerank). This is because using 1x1 filters only may not be easy to capture some information such as the variance of values in a column or row. 
Second, we replace all filters with filters of size 5x5 that is common in computer vision tasks~\cite{alexnet, vgg}, which leads to another substantial drop. Both results confirm the effectiveness of our model design that is tailored to table tasks.

\stitle{Sensitivity analysis}
We perform sensitivity analysis to understand the effect of different settings in \at.

\underline{Varying Input Size.} In \at, we feed the top 100 rows and left-most 50 columns from the input table $T$ into the model, which is typically enough to correctly predict the right transformations. To understand its effect on model performance, in Figure ~\ref{fig:vary_input}, we vary the number of rows/columns used here and show the input-only model performance. As we can see, when we increase the number of rows/columns that the model uses, the resulting quality improves until it plateaus at about 30 columns and 50 rows.% input rows and columns. Also, we can that it is more sensitive to change of column size than row size. However, as the the number of rows and columns becomes large enough, the improvement becomes insignificant. 

% \textcolor{red}{add latency}

\underline{Varying Number of Filters.} Figure~\ref{fig:vary_filter} shows the quality of \at input-only model with different numbers of convolution filters (the total number of 1x1 and 1x2 filters for rows/columns before AvgPool in the feature extraction layer in Figure~\ref{fig:input_model_arch}). As we can see, using 32 filters is substantially better than 4 filters, as it can extract more features. However, the improvement beyond 32 filters is not significant, suggesting diminishing returns beyond a certain level of model capacity.

% using 32 filters is substantially better than 4 filters, suggesting more than a few latent ``patterns'' in table data. The quality of our results then plateaus, as using 64 filters does not improve the results.

% removed for revision, add back
%\revised{}
%\begin{comment}
\iftoggle{fullversion}
{
    \underline{Varying Embedding Methods. } We initially choose the powerful (but expensive) sentenceBERT~\cite{reimers-2019-sentence-bert} as our semantic embedding, which is known to excel in NLP tasks. We explore how alternative embeddings, such as GloVe~\cite{pennington2014glove}, and fastText~\cite{bojanowski2017enriching}, would perform in our task. %, considering the fact that sentenceBERT is the most expensive part in \at in terms of latency (Figure~\ref{fig:running_time}).
    We show the performance of input-only model with different embeddings in Table~\ref{tab:vary_embed}. As we can see, \at is interestingly not sensitive to the exact choice of semantic embedding --  using sentenceBERT/GloVe/fastText achieves similar quality, suggesting that \at  can operate at a much lower latency than was shown in Figure~\ref{fig:running_time}, without loss of quality.
    %\end{comment}
}
{

}

% removed for revision, add back
%\revised{}
%\begin{comment}
%\subsection{Error Analysis.} 

\iftoggle{fullversion}
{
    \stitle{Error Analysis.} 
    We analyze mistakes that the \at model makes on 218 tables that need a single-step transformation. We show the errors in both predicting operator-type and parameters.
    
    Table~\ref{tab:confusion_matrix} shows a detailed confusion matrix for single-step top-1 operator-type predictions. We can see that the most common mistakes are between ``\code{transpose}'' and ``\code{stack}'' (9), as well as ``\code{wide-to-long}'' and ``\code{stack}'' (6). Both are not unexpected, as their corresponding input tables share similar characteristics (e.g., the input in Figure~\ref{fig:multi-step-ex} may appear to look like a candidate for ``\code{transpose}'' as well as ``\code{stack}'', due to its homogeneous column groups).
    
    %of different methods, where we only run each method with one step and only evaluate it on tables that need one-step transformation. Note that this task is easier than end-to-end task, which can involve multiple operators and must stop by predicting None correctly when the input table is already relationalized. As we can see, \at performs better than all other methods by a large margin. 
    
    Table~\ref{tab:parameters} shows the accuracy of our parameter predictions for different operators at the top-1 position. % (these parameters are described in Table~\ref{tab:label_vector_encode}). 
    Despite the large space of possible parameters, our predictions are surprisingly accurate, showing the effectiveness of our CNN-inspired model in extracting patterns from tabular data.

    % \begin{table}[!h]
    % \caption{Confusion matrix for single-step top-1 predictions.}
    % \vspace{-4mm}
    % \scalebox{0.7}{
    % \begin{tabular}{|c|*{8}{c|}}\hline
    % \backslashbox{\textbf{True}}{\textbf{Pred}}
    % &\makebox[2em]{\textbf{trans.}}&\makebox[2em]{\textbf{stack}}
    % &\makebox[2em]{\textbf{wtl}}&\makebox[3em]{\textbf{explode}}&\makebox[2em]{\textbf{ffill}}&\makebox[2em]{\textbf{pivot}}&\makebox[3em]{\textbf{subtitle}}&\makebox[2em]{\textbf{none}}\\\hline
    % \textbf{trans.} & 12 & 9 & 0 & 1 & 0 & 0 & 0 & 0 \\ \hline
    % \textbf{stack} & 2 & 24 & 1 & 0 & 0 & 0 & 0 & 8 \\ \hline
    % \textbf{wtl} & 0 & 6 & 21 & 0 & 0 & 0 & 0 & 4 \\ \hline 
    % \textbf{explode} & 0 & 0 & 0 & 19 & 0 & 0 & 0 & 14 \\ \hline
    % \textbf{ffill} & 0 & 1 & 0 & 1 & 8 & 0 & 0 & 3 \\ \hline
    % \textbf{pivot} & 0 & 0 & 0 & 0 & 0 & 8 & 0 & 0 \\ \hline
    % \textbf{subtitle} & 0 & 1 & 0 & 0 & 0 & 0 & 24 & 1 \\ \hline
    % \end{tabular}
    % }
    % \label{tab:confusion_matrix}
    % \end{table}
    
    \begin{table}[!h]
    \caption{Confusion matrix for single-step top-1 predictions.}
    %\vspace{-4mm}
    \scalebox{0.7}{
    \begin{tabular}{|c|*{8}{c|}}\hline
    \backslashbox{\textbf{True}}{\textbf{Pred}}
    &\makebox[2em]{\textbf{trans.}}&\makebox[2em]{\textbf{stack}}
    &\makebox[2em]{\textbf{wtl}}&\makebox[3em]{\textbf{explode}}&\makebox[2em]{\textbf{ffill}}&\makebox[2em]{\textbf{pivot}}&\makebox[3em]{\textbf{subtitle}}&\makebox[2em]{\textbf{none}}\\\hline
    \textbf{trans.} & 14 & 10 & 1 & 1 & 0 & 0 & 0 & 2 \\ \hline
    \textbf{stack} & 2 & 36 & 3 & 1 & 0 & 0 & 0 & 14 \\ \hline
    \textbf{wtl} & 0 & 6 & 23 & 1 & 0 & 0 & 0 & 4 \\ \hline 
    \textbf{explode} & 0 & 0 & 0 & 32 & 0 & 0 & 0 & 16 \\ \hline
    \textbf{ffill} & 0 & 1 & 0 & 1 & 10 & 0 & 0 & 6 \\ \hline
    \textbf{pivot} & 0 & 0 & 0 & 0 & 0 & 8 & 0 & 0 \\ \hline
    \textbf{subtitle} & 0 & 1 & 0 & 0 & 0 & 0 & 24 & 1 \\ \hline
    \end{tabular}
    }
    \label{tab:confusion_matrix}
    \end{table}

    % \begin{table}[!h]
    % \caption{Accuracy of  operator parameter predictions}
    % \vspace{-4mm}
    % \scalebox{0.68}{
    % \begin{tabular}{cccccccc}
    % \toprule
    % %\textbf{Parameters} & \textbf{ssi} & \textbf{sei} & \textbf{wsi} & \textbf{wei} & \textbf{eci} & \textbf{fei} & \textbf{prf} \\
    % \textbf{operator} & \textbf{stack} & \textbf{stack} & \textbf{wtl} & \textbf{wtl} & \textbf{explode} & \textbf{ffill} & \textbf{pivot} \\
    % \textbf{parameter} & \textbf{start-idx} & \textbf{end-idx} & \textbf{start-idx} & \textbf{end-idx} & \textbf{col-idx} & \textbf{col-idx} & \textbf{row-freq} \\
    % \midrule
    % Accuracy & 0.958 & 1 & 0.952 & 1 & 0.947 & 1 & 0.875 \\
    % \bottomrule
    % \end{tabular}
    % }
    % \label{tab:parameters}
    % \end{table}
    
    \begin{table}[!h]
    \caption{Accuracy of  operator parameter predictions}
    %\vspace{-4mm}
    \scalebox{0.68}{
    \begin{tabular}{cccccccc}
    \toprule
    %\textbf{Parameters} & \textbf{ssi} & \textbf{sei} & \textbf{wsi} & \textbf{wei} & \textbf{eci} & \textbf{fei} & \textbf{prf} \\
    \textbf{operator} & \textbf{stack} & \textbf{stack} & \textbf{wtl} & \textbf{wtl} & \textbf{explode} & \textbf{ffill} & \textbf{pivot} \\
    \textbf{parameter} & \textbf{start-idx} & \textbf{end-idx} & \textbf{start-idx} & \textbf{end-idx} & \textbf{col-idx} & \textbf{col-idx} & \textbf{row-freq} \\
    \midrule
    Accuracy & 0.889 & 1 & 0.957 & 1 & 0.969 & 1 & 0.875 \\
    \bottomrule
    \end{tabular}
    }
    \label{tab:parameters}
    \end{table}
    %\end{comment}
}
{
\underline{Additional results.} We report  additional results such as sensitivity to different embeddings, error analysis, and accuracy of parameter predictions, in~\cite{full} in the interest of space.
}

%\stitle{\yeye{Varying number of layers?}}
% \yeye{after reading the experiments, I have a feeling that some reviewers may think we don't have enough experiments yet (we have 1 table and 2 figures right now). Maybe we can add some  sensitivity analysis? E.g., (1) sensitivity to trainng corpus (pbi, parquet, pbi+parquet?), (2) sensitivity to hyperparameters, e.g., number of filters, etc. (3) others?}

%% file: Conclusions.tex
\section{Conclusions and Future Work}
We propose a new problem of synthesizing transformations to relationalize tables. By leveraging visual characteristics of input tables using compute-vision-inspired algorithms, we obviate the need for users to provide input/output examples, which is a substantial departure from prior work. 
Future directions include extending the functionality to a broader set of operators, and exploring the applicability of this technique on other classes of transformations.